\documentclass[ps,prd,preprint,
superscriptaddress,preprintnumbers,eqsecnum,
showpacs,nofootinbib,nobibnotes]{revtex4}
\usepackage{amsfonts,bm}
\usepackage{graphicx}
\usepackage{pstricks}
\usepackage{color}

\newcommand{\be}{\begin{equation}}
\newcommand{\bea}{\begin{eqnarray}}
\newcommand{\ee}{\end{equation}}
\newcommand{\eea}{\end{eqnarray}}

\def\ope{\omega}
\def\operhs{\Upsilon}
\def\homo{\kappa}
\def\Y{\Xi}

\def\diff{\mathrm{d}}

\def\fg{\mathrm{I}\!\Gamma}
\def\G{\Gamma}

\def\rG{\widetilde{\G}}

\def\g{{\cal G}}
\def\zb{(A,\widehat{A})=0}

\def\aeta{\bar \eta}

\begin{document}

\title{Slavnov-Taylor constraints for non-trivial backgrounds}
\date{June 16, 2011}

\author{D. Binosi}
\email{binosi@ect.it}
\affiliation{European Centre for Theoretical Studies in Nuclear
  Physics and Related Areas (ECT*) and Fondazione Bruno Kessler, \\ Villa Tambosi, Strada delle
  Tabarelle 286, 
 I-38123 Villazzano (TN)  Italy}
 
\author{A. Quadri}
\email{andrea.quadri@mi.infn.it}
\affiliation{Dip. di Fisica, Universit\`a degli Studi di Milano via Celoria 16, I-20133 Milano, Italy\\
and INFN, Sezione di Milano, via Celoria 16, I-20133 Milano, Italy}

\begin{abstract}
\noindent
We devise an algebraic procedure for the
evaluation of Green's functions in $SU(N)$
Yang-Mills theory in the presence of a 
non-trivial background field. 
In the ghost-free sector the dependence
of the vertex functional  on the background
is shown to be uniquely determined by the 
Slavnov-Taylor identities in terms of
a certain 1-PI
correlator of the covariant derivatives
of the ghost and the anti-ghost fields.
At non-vanishing background this amplitude is shown to encode the
quantum deformations to the tree-level
background-quantum splitting.
The approach only relies on the
functional identities of the model
(Slavnov-Taylor identities, $b$-equation, 
anti-ghost equation)
and thus it is valid beyond perturbation theory, and in particular in a lattice implementation of the background field method.
As an example of the formalism we analyze the
ghost two-point function and the Kugo-Ojima function 
in an instanton background 
in $SU(2)$ Yang-Mills theory,
quantized in the background Landau gauge.
\end{abstract}

\pacs{
11.15.Tk,	
12.38.Aw,  
12.38.Lg
}

\maketitle

\section{Introduction}

\noindent The Background Field Method (BFM)~\cite{DeWitt:1967ub,Abbott:1980hw} is known to be a very powerful tool for studying the properties of non-Abelian gauge theories. Its main advantage over the conventional quantization formalisms ({\it e.g.}, the  ordinary renormalizable~$R_\xi$ gauges) resides in the fact that it preserves gauge invariance with respect to the background field at the quantum level, thus providing a linear functional identity (the background Ward identity) for the vertex functional. This identity leads in turn to linear relations between the 1-PI Green's functions of the theory, as opposed to the bilinear relations obtained from the Slavnov-Taylor identities.
It then follows that the construction of the (background) effective action is significantly simplified, for the structure of the allowed counterterms is  greatly constrained by the symmetry requirements enforced by the background gauge invariance. This, together with its perturbative equivalence with the usual perturbation theory based on the Gell-Mann and Low's formula~\cite{Becchi:1999ir,Ferrari:2000yp}, makes the application of the BFM advantageous in a variety of situations, ranging from perturbative calculations in Yang-Mills theories~\cite{Abbott:1980hw,Ichinose:1981uw} via the quantization of the Standard Model~\cite{Denner:1994xt} to gravity and supergravity calculations~\cite{Gates:1983nr}.

Another important aspect of the BFM is that it can be used as a simple prescription for calculating to any order the $n$-point Green's functions of the pinch technique~\cite{Cornwall:1981zr, Cornwall:1989gv}. This technique represents the only known method capable of enforcing explicit gauge invariance in (all-order) off-shell Green's functions~\cite{Binosi:2002ft} and the corresponding infinite set of  (non-perturbative) integral equations (the so-called Schwinger-Dyson equations) that couples them~\cite{Binosi:2007pi}. Specifically, this equivalence stems from an infinite tower of powerful identities -- the so-called {\it background-quantum} identities~\cite{Grassi:1999tp,Binosi:2002ez} -- that relates Green's functions involving a given combination of quantum and background fields with the same  functions where one of the background fields has been replaced by its quantum counterpart.

Indeed, these identities play a fundamental role in the two-point sector of (pure) $SU(N)$ Yang-Mills theories, where it is known that the IR behavior of the gluon (and ghost) propagator encodes precious information about the non-perturbative dynamics of the theory in general, and the confinement phenomenon in particular. In this case in fact one can study the Schwinger-Dyson equation for the background ({\it viz.} the pinch technique) gluon propagator, which can  be  truncated gauge invariantly by exploiting the block-wise transversality of its gluon and ghost one- and two-loop dressed contributions~\cite{Binosi:2007pi,Aguilar:2006gr}. The solution of this equation can be then related to the conventional one through the corresponding two-point background quantum identity; the result is {\it a gauge artifact free propagator} that can be meaningfully compared to the plethora of high quality {\it ab-initio} lattice gauge theory computations currently available~\cite{Aguilar:2008xm}.

The combination of the aforementioned continuum studies and lattice data gives overwhelming evidence that (in the Landau gauge) the gluon propagator and the ghost dressing function saturates in the deep IR at a finite, non-vanishing value~\cite{Cucchieri:2007md,Bogolubsky:2007ud}. The lattice preference for these so-called {\it massive solutions} has entailed a paradigmatic shift in our understanding of the QCD IR dynamics, thus forcing  the abandoning of the original formulation of confinement scenarios such as the ones of Kugo-Ojima~\cite{Kugo:1979gm} (predicting an IR divergent -- or {\it enhanced} -- ghost dressing function) and Gribov-Zwanziger~\cite{Gribov:1977wm,Zwanziger:1993dh} (predicting an IR divergent ghost dressing function and an IR vanishing gluon propagator) in favor of models capable of accommodating a dynamically generated gluon mass~\cite{Cornwall:1981zr,Dudal:2008sp}. 

Possible loopholes in lattice studies have been meanwhile also thoroughly addressed. In fact, when calculating off-shell Green's functions on the lattice one does not only need to fix a gauge, but has also to keep under control the various sources of systematic errors (e.g., discretization effects, finite volume effects, Gribov copies effects), while at the same time providing enough computational power (read large volume lattices) to study the deep IR region~\cite{Cucchieri:2010xr}. Of all these problems, the most pressing and debated one is probably the Landau gauge projection, which is well-known to suffer from  Gribov copies; yet there are clear indications that the effects of such copies is quantitative (rather than qualitative) and well under control~\cite{Cucchieri:2010xr}.  

Thus, given the state-of-the-art just described, it would be highly desirable to compute the gluon (and ghost) propagator and study their IR behavior in as many gauges  as possible. Progress in implementing the $R_\xi$ gauges for $\xi\neq0$, and in particular the Feynman gauge, has been recently reported~\cite{Cucchieri:2011aa}; nonetheless, it is clear that implementing the BFM method on the lattice (for whatever value of the gauge fixing parameter) would be a long awaited leap forward~\cite{Dashen:1980vm}. 

Putting the BFM on a lattice requires the choice of a suitable background field $\widehat{A}$. This is a more subtle operation than it looks like at a first sight. To understand why, let's concentrate to the background Landau gauge case, which is fixed by the condition $\widehat{\cal D}^\mu Q_\mu=0$, with $\widehat{\cal D}$ the background covariant derivative and $Q=A-\widehat{A}$ the quantum field -- see Section~\ref{prel} below for our notation. Then, it has been shown in~\cite{Zwanziger:1982na} that the gauge can be fixed locally if and only if given an infinitesimal gauge transformation
of parameter $w$ in the Lie algebra of the gauge group
such that $A_\mu\to A_\mu+{\cal D}_\mu w$ (${\cal D}$ being the ordinary covariant derivative, see Section~\ref{prel} again), the equation
\be
\widehat{\cal D}^\mu{\cal D}_\mu w=0
\ee
has no solutions other than $w=0$. Since in the background Landau gauge the background and covariant derivatives commute, this implies that the latter condition does not fix uniquely the gauge (not even locally) if the equation
\be
\widehat{\cal D}_\mu w=0,
\label{npf}
\ee
has any solution $w\neq 0$. If a solution of the latter type exists, the background gauge potential $\widehat{A}$ is called {\it partially flat}; a good background on the lattice is therefore non partially flat~\cite{Zwanziger:1982na}. Notice that, in a finite volume system as the lattice, this rules out the naive (perturbative) vacuum $\widehat{A}=0$, otherwise $w=w_0$ with $w_0$ constant, would be an acceptable solution of~(\ref{npf});
on the other hand, the latter is precisely the standard vacuum used in the analyses carried out in the literature when attempting to solve the constraints coming from the defining functional identities of the theory (Slavnov-Taylor, background-quantum, $b$-equation, ghost and anti-ghost equations).

The present paper serves precisely the purpose of developing the new formal tools needed to solve the relevant functional identities in those cases where one has to deal with a non-trivial (or non partially flat in the lattice case) background configurations, {\it e.g.,} the topologically non-trivial vacuum configurations provided by vortices, monopoles and  instantons. The upshot of our analysis will be that Green's functions involving only gluon fields in a non-trivial background can be obtained from the evaluation of the same amplitudes at zero background field, once one performs a gluon field redefinition which generalizes the quantum-background replacement $Q=A-\widehat{A}$ when loop corrections are taken into account. The latter field redefinition can be explicitly
computed in terms of a certain functional, involving the insertion of two composite operators given by the BRST variation of the gluon field and the covariant derivative of the antighost.

This opens also up the possibility of encoding topological information (such as winding numbers) into  continuum non-perturbative methods ({\it e.g.}, the aforementioned Schwinger-Dyson equations) by calculating the correction terms due to the presence of a non-trivial background $\widehat{A}\neq0$.  In this way one might be able to describe what happens when topological effects are properly taken into accounts, and compare with what has been observed on the lattice when center vortices are removed from the vacuum configurations~\cite{de Forcrand:1999ms,Gattnar:2004bf}.

The paper is organized as follows.
Since our results will be derived within the quantization framework of Batalin and Vilkovisky~\cite{Batalin:1977pb}, we start by briefly recalling its main ingredients in Section~\ref{prel}. 
Unlike in the conventional BFM formalism, we keep the background
as a fixed classical non-trivial configuration. 
This entails that we do not rely
on the background equivalence theorem \cite{Ferrari:2000yp}, 
which allows in perturbation theory 
to derive the connected amplitudes of gauge-invariant physical operators
by taking the Legendre transform with respect to the background fields (and not
with respect to the quantized fields, as is prescribed by the Gell-Mann and Low's formula).
Next, we will analyze the consequences of allowing a non-trivial background by looking in detail at the two-point ghost sector. We will then move to the central result of the paper that is the determination of the functional encoding  the deformation of the background-quantum splitting induced by quantum corrections (Section IV) and the complete solution of the recursion for the background amplitudes (Section V). After drawing our conclusions and look into possible applications of the results presented, in the Appendix some perturbative results and checks will be discussed.

\section{\label{prel}Preliminaries}

\noindent When dealing with theories possessing a non-linear BRST operator $s$, such as $SU(N)$ Yang-Mills theories in general, and QCD in particular, an efficient procedure to quantize the theory is through the introduction of certain external sources $\phi^*$ (one for each field $\phi$ transforming non-linearly under $s$)  describing the renormalization of the composite operators that are bound to appear.  These sources, called {\it anti-fields}~\cite{Gomis:1994he}, have opposite statistics with respect to the corresponding field $\phi$, ghost charge $\mathrm{gh}(\phi^*)=-1-\mathrm{gh}(\phi)$, and,  choosing the (mass) dimension of the Faddeev-Popov ghost fields to be zero, dimension $\mathrm{dim}(\phi^*)=4-\mathrm{dim}(\phi)$ (see Table~\ref{tableI}).
This ensures that the Lagrangian ${\cal L}$ has ghost number zero and canonical dimension four.

\begin{table}
\begin{center}
\begin{tabular}{r||c|c|c|c|c|c||c|c|}
 & $\ A^a_\mu\ $ &  $\ c^a\ $ & $\ \bar c^a\ $  & $\ b^a\ $ &  $\ A^{*a}_\mu\ $ & $\ c^{*a}\ $  & $\ \widehat{A}^a_\mu\ $ & $\ \Omega^a_\mu\ $\\
\hline\hline
Ghost charge & 0  & 1 & -1  & 0  & -1  & -2 & 0 & 1\\
\hline
Statistics  & B & F & F  & B & F &  B & B & F\\
\hline
Dimension & 1 &  0 & 2 & 2 & 3 &  4  & 1 & 1 \\
\hline
\end{tabular} 
\caption{Ghost charge, statistics (B for Bose, F for Fermi), and mass dimension of both the $SU(N)$ Yang-Mills conventional fields and anti-fields as well as background fields and sources. \label{tableI}}
\end{center}
\end{table}

Anti-fields are then coupled in the tree-level vertex functional
$\G^{(0)} = \int\diff^4x \, {\cal L}$ to the quantum fields through the term  $\sum\phi^*s\, \phi$, where, specializing to $SU(N)$ Yang-Mills theories and neglecting matter fermion fields, one has
\bea
s A^a_\mu=({\cal D}_\mu c)^a; &\quad& sc^a=-\frac12f^{abc}c^bc^c\nonumber \\
s\bar c^a=b^a;&\quad&sb^a=0.
\label{brst.s}
\eea
In the expressions above ${\cal D}$ represents the covariant derivative 
in the adjoint representation of the gauge group, {\it i.e.},
\be
({\cal D}_\mu\phi)^a={\cal D}^{ab}_\mu\phi^b;\qquad 
{\cal D}^{ab}_\mu=\delta^{ab}\partial _\mu +f^{acb}A^c_\mu
\ee
while the $b$ field is the Nakanishi-Lautrup multiplier for the (yet to be specified) gauge fixing 
function  ${\cal F}$, so that the gauge fixing and Faddeev-Popov ghost Lagrangian will be given by the total BRST variation
\be
{\cal L}_{{\mathrm{GF}}}+{\cal L}_{{\mathrm{FPG}}}=s\left(\bar c^a\mathcal{F}^a-\frac\xi2\bar c^a b^a\right).
\ee
The tree-level vertex functional is then written as
\be
\fg^{(0)}=\int\!\mathrm{d}^4x\left[-\frac14F^a_{\mu\nu} F^{a\mu\nu}+{\cal L}_{{\mathrm{GF}}}+{\cal L}_{{\mathrm{FPG}}}+A^{*a}_\mu\left({\cal D}^\mu c\right)^a-\frac12f^{abc} c^{*a}c^bc^c\right].
\label{tlvf}
\ee
 
In order to specialize to the BFM type of gauges, which represents the relevant case for the ensuing analysis, let us split the classical field $A$ into a background ($\widehat{A}$) and a quantum ($Q$)  part according to
\be
A_\mu^a=\widehat{A}_\mu^a+Q_\mu^a.
\ee 
Next, we retain the background gauge invariance 
of the gauge-fixed action by choosing a gauge-fixing
function transforming in the adjoint representation of $SU(N)$
through the replacement of the ordinary derivative with
the background covariant derivative
\bea
{\cal F}^a&=&(\widehat{{\cal D}}^\mu Q_\mu)^a\nonumber \\
&=& \partial^\mu Q_\mu^a+f^{abc}\widehat{A}^b_\mu Q^c_\nu.
\eea
As a last step, in addition to the anti-fields $\phi^*$, the quantization of the theory in the BFM requires the introduction of an additional (vector) source $\Omega$, implementing at the quantum level the equation of motion of the background field $\widehat{A}$, with
\be
s\widehat{A}^a_\mu=\Omega^a_\mu; \qquad s\Omega^a_\mu=0.
\label{bkg.brst}
\ee

In ordinary perturbative quantum field theory, Eq.~(\ref{bkg.brst}) implements the so-called doublet mechanism~\cite{Piguet:1995er,Barnich:2000zw,Quadri:2002nh},  preventing the background field from modifying the physical observables of the model. Briefly, a pair of variables $(u,v)$ such that $su=v$, $s v =0$ is called a BRST doublet ($v$ represents the BRST partner of $u$). In the BRST quantization
approach, the physical observables ({\it i.e.}, the set of physical local operators)
admit a mathematical characterization in terms of the local cohomology
$H^0(s)$ of the BRST operator $s$ in ghost number zero~\cite{Barnich:2000zw,Barnich:1994ve}.
The latter is defined by identifying all the local zero ghost number operators in the kernel of $s$ which differ by a total $s$-variation; that is, we say that two BRST-invariant operators ${\cal O}(x)$ and ${\cal O}'(x)$ are equivalent iff they can be written as 
\be
{\cal O}(x) = {\cal O}'(x) + s {\cal Q}(x)
\label{cohom.1}
\ee
for some local operator ${\cal Q}(x)$ with ghost number $-1$.
One is also interested in the cohomology $H^0(s|d)$
of the BRST differential $s$ modulo the
exterior derivative $d$. This is given by BRST-invariant integrated
local operators with ghost number zero when one
identifies operators
differing by a total $s$-variation.
For $SU(N)$ Yang-Mills theories with no matter fermions $H^0(s|d)$ is given by all integrated gauge-invariant
polynomials constructed out of the field strength $F^a_{\mu\nu}$ for the gauge field $A$ and its
covariant derivatives \cite{Barnich:2000zw,Barnich:1994ve}.
Notice that the latter do not depend on the background field and on its BRST partner, as
a consequence of a general theorem \cite{Barnich:2000zw,Quadri:2002nh} stating that doublet variables drop out in the computation of the cohomology of the BRST differential.

On the other hand, if one considers, as we do here, the computation of gauge-variant quantities like {\it e.g.}, Green's functions of the ghost fields, there is no reason to exclude
a (non perturbative) dependence on a non-trivial background configuration.

The BRST transformation of the quantum field $Q$ is from Eq.~(\ref{bkg.brst}) and
Eq.~(\ref{brst.s})
\be
sQ^a_\mu=sA^a_\mu-\Omega^a_\mu=\fg^{(0)}_{A^{*a}_\mu}-\Omega^a_\mu,
\ee
where, for later convenience, we have introduced the notation
$\G_\phi\equiv\delta_\phi\G\equiv\frac{\delta}{\delta\phi}\G$ with $\G$ an arbitrary functional of $\phi$.
One is then led to the Slavnov-Taylor (ST) identity in functional form~\cite{Grassi:1999tp} 
\be
\int\!\mathrm{d}^4x\left[\fg_{A^{*\mu}_a}\fg_{Q^{a}_\mu}+\fg_{c^{*a}}\fg_{c^a}+b^a\fg_{\bar c^a}+\Omega^\mu_a\left(
\fg_{\widehat{A}^a_\mu}-\fg_{Q^a_\mu}\right)\right]=0 \, ,
%
\label{me}
\ee
where $\fg$ is now the (quantum) effective action. Notice that the ST identity  above can also be rewritten in terms of the original field $A$ to assume the somewhat more compact form
\be
\int\!\mathrm{d}^4x\left[\fg_{A^{*\mu}_a}\fg_{A^{a}_\mu}+\fg_{c^{*a}}\fg_{c^a}+b^a\fg_{\bar c^a}+\Omega^\mu_a
\fg_{\widehat{A}^a_\mu}\right]=0.
\label{me-1}
\ee
By setting the background field and source to zero one recovers the usual ST identity in the ordinary $R_\xi$ gauges.

The usual Slavnov-Taylor identities are  generated from Eq.~(\ref{me}) -- or Eq.~(\ref{me-1}) -- by taking functional differentiations with respect to combinations of fields containing either one ghost field, or two ghost fields and one anti-field, setting all fields/sources to zero afterwards (the only exception to this rule being when differentiating with respect to a ghost anti-field, which needs to be compensated by three ghost fields). In contrast, functional differentiation with respect to a background source and background and/or quantum fields will provide the so-called background-quantum identities which relate Green's functions involving background fields to those involving quantum fields~\cite{Grassi:1999tp,Binosi:2002ez}.

A further Ward-Takahashi identity holds in the background
gauge as a consequence of the invariance under background
gauge transformations:
\be
{\cal W}_a(\fg) = -\partial^\mu \frac{\delta \fg}{\delta \widehat{A}^a_\mu}
+  f^{abc} \widehat{A}^\mu_c \frac{\delta \fg}{\delta \widehat{A}^b_\mu}
+ \sum_{\chi \in \{ Q,c,\bar c,b,\Omega,\phi^* \}}
 f^{abc} \chi^c \frac{\delta \fg}{\delta \chi^b} = 0  .
\label{wti}
\ee
Notice that this identity is linear in the vertex functional,
unlike the ST identity (\ref{me-1}).

The linearity  of the gauge fixing function in the quantum fields implies also the existence of a constraint coming from the equation of motion of the $b$ field
\be
\fg_{b^a}=-\xi b^a + (\widehat{\cal D}^\mu Q_\mu)^a,
\label{beq}
\ee
which takes the form of the ghost (or Faddeev-Popov) equation
\be
\fg_{\bar c^a}+(\widehat{\cal D}^\mu\fg_{A^*_\mu})^a-\left({\cal D}^\mu\Omega_\mu\right)^a=0.
\label{FPE}
\ee

Finally, when considering the background Landau gauge $\xi=0$, one has
\bea
{\cal L}_{{\mathrm{GF}}}+{\cal L}_{{\mathrm{FPG}}}&=&s\left[\bar{c}^a(\widehat{\cal D}^\mu Q_\mu)^a\right]\nonumber \\
&=&b^a(\widehat{\cal D}^\mu Q_\mu)^a-\bar c^a\left[(\widehat{\cal D}^\mu{\cal D}_\mu c)^a-(\widehat{\cal D}^\mu\Omega_\mu)^a+ f^{abc}\Omega_b^\mu
Q^c_\mu\right].
\eea
As a consequence, an additional equation appears namely the anti-ghost equation~\cite{Grassi:2004yq}
\be
\fg_{c^a}-(\widehat{\cal D}^\mu\fg_{\Omega_\mu})^a+f^{abc}\fg_{b^b}\bar{c}^c
-\left({\cal D}^\mu A^*_\mu \right)^a
-f^{abc}c^{*b}c^c=0.
\label{AGE}
\ee
Notice that this equation is local, as opposed to the integrated (and correspondingly less powerful) equation one would get within the Landau gauge in the conventional $R_\xi$ gauges.

In addition, in this gauge the background Ward-Takahashi identity~(\ref{wti}) is not an independent identity; in fact, introducing the linearized ST operator ${\cal S}_{\fg}$ acting on a functional $X$ as
\be
{\cal S}_{\fg}(X) = \int\!\diff^4x \left [ 
\fg_{A^{*\mu}_a}X_{A^{a}_\mu}+\fg_{A^{a}_\mu} X_{A^{*\mu}_a}+\fg_{c^{*a}}X_{c^a}
+\fg_{c^a}X_{c^{*a}}+b^aX_{\bar c_a}
+\Omega^\mu_a X_{\widehat{A}^a_\mu}
\right],
\ee
Eq.~(\ref{wti}) can be rewritten as the anticommutator of the ST identity
(\ref{me-1}) and the antighost equation (\ref{AGE}), that is
\bea
{\cal W}^a(\fg) &=&{\cal S}_{\fg} \Big (\fg_{c^a}-(\widehat{\cal D}^\mu\fg_{\Omega_\mu})^a+gf^{abc}\fg_{b^b}\bar{c}^c
-\left({\cal D}^\mu A^*_\mu \right)^a
-f^{abc}c^{*b}c^c \Big ) + \frac{\delta}{\delta c^a}
{\cal S}(\fg)\nonumber\\&=&  0  .
\eea

\section{Two-point functions in a non-trivial background}\label{sec:2point.id}

In this section we start looking into the consequences of allowing a non-trivial background by considering as a case study the two-point ghost sector; in particular we will highlight the differences induced in the definition of the 1-PI functions and the corresponding relations dictated by the functional identities~(\ref{me-1}), (\ref{FPE}) and~(\ref{AGE}). 

Let us begin by introducing the following definition of the 1-PI Green's functions $\G_{\phi_1 \dots \phi_n \phi^*_1 \dots \phi^*_m}$
\be
\G_{\phi_1 \dots \phi_n \phi^*_1 \dots \phi^*_m} = \left. \frac{\delta^{(n+m)} \fg}{\delta \phi_1 \dots \delta \phi_n
\delta \phi^*_1 \dots \delta \phi^*_m
} \right |_{\widehat{A} \neq 0}\hspace{-.78cm}{\scriptsize \matrix{\vspace{-.2cm}(\phi^*,c,\bar c,\Omega)=0}},
\label{not-I}
\ee
where $\phi$ includes now background gluons $\widehat{A}$ as well, and 
the indices $1, \dots, n, \dots m$ denote the dependence on the internal and Lorentz indices as well as on coordinates.
Thus the Green's functions $\G_{\phi_1 \dots \phi_n \phi^*_1 \dots \phi^*_m}$ are calculated by setting all fields and external sources but $\widehat{A}$ to zero, since we
want eventually to compute them  in a non-perturbative setting where
non-trivial background $\widehat{A}\neq0$ is present; we will also set $\G = \left . \fg \right |_{\Omega=0}$.

Next, we differentiate Eq.~(\ref{AGE}) with respect to $\bar c_b$; then, by using the $b$-equation (\ref{beq}) we find
\be
\G_{\bar c^b c^a} = {\widehat {\cal D}}_\mu^{ac} \G_{\bar c^b \Omega_c^\mu} + f^{acb} \partial^\mu \widehat{A}^c_\mu\delta(x-y). 
\label{sec.2.1}
\ee
In a similar fashion the differentiation of Eq.~(\ref{AGE}) with respect to $A^{*b}_\mu$ yields
\be
\G_{A^{*b}_\mu c^a} = \delta^{ab} \partial_\mu  \delta(x-y)
- {\widehat{\cal D}}_\nu^{ac} \G_{\Omega_c^\nu A^{*b}_\mu}.
\label{age.2}
\ee
Finally, differentiating Eq.~(\ref{FPE})  with respect to $\Omega^b_\mu$ we get
\be
\G_{\Omega^b_\mu \bar c^a} =
-{\widehat{\cal D}}_\rho^{ac} \G_{\Omega^b_\mu A^{*\rho}_ c}
+ \partial_\mu \delta(x-y) \delta^{ba}.
\label{sec.2.2}
\ee
By substituting Eq.~(\ref{sec.2.2}) into Eq.~(\ref{sec.2.1}) we arrive at the final answer
\be
\G_{\bar c^b c^a}  = 
 \square \delta^{ab} \delta(x-y) 
+  f^{abc} \widehat{A}^c_\mu \partial^\mu \delta(x-y) - f^{abc} \partial^\mu {\widehat A}^c_\mu \delta(x-y) 
+  {\widehat{\cal D}}_\mu^{ac} 
{\widehat {\cal D}}_\nu^{bd}
\G_{\Omega^\mu_c A^{*\nu}_d}.
\label{sec.2.3}
\ee
Thus the two-point 1-PI ghost Green's function is fully
determined in the background Landau gauge 
by the Green's function $\G_{\Omega A^*}$ {\it alone}, even in the presence of a non-trivial background configuration.

The latter function can be explored by non-perturbative methods ({\it e.g.} evaluating it on the lattice by means of Monte Carlo averages) through the connected Green's function
\be
{\cal C}_{\mu\nu}^{ab}(x,y) = 
\langle T[ ({\cal D}_\mu \bar c)^a (x)({\cal D}_\nu c)^b (y)]\rangle^C
= \left. \frac{\delta^2 W}{\delta \Omega^a_\mu(x) 
\delta A^{*b}_\nu(y)} \right |
_{\widehat{A} \neq 0}\hspace{-.78cm}{\scriptsize \matrix{\vspace{-.2cm}(J,\phi^*, \Omega)=0}},
\label{sec.3.n.1}
\ee
where $T$ indicates the time ordered product of fields, and 
$W$ is the connected generating functional, obtained
by taking the Legendre transform of $\fg$ w.r.t
to $\phi$
\be
W = \fg + \int\!\diff^4x \, J\cdot\phi
\label{wfunct}
\ee
($J$ is a collective notation for the sources of the
quantized fields $\phi$).

In fact the function ${\cal G}$ can be decomposed into
its connected components according to
\be
{\cal C}_{\mu\nu}^{ab}(x,y) = 
\G_{\Omega^a_\mu A^{*b}_\nu}(x,y) + 
{ i}
\int\!\diff^4z\!\!\int\!\diff^4w \, 
\G_{\Omega^a_\mu \bar c^{d}}(x,z)
D^{d e}(z,w)
\G_{c^{e} A^{*b}_\nu}(w,y),
\label{Cconn}
\ee
where $D$ denotes the dressed ghost propagator obtained
by inverting Eq.~(\ref{sec.2.3}).

Use of Eqs.~(\ref{age.2}) and~(\ref{sec.2.2}) yields
\begin{eqnarray}
{\cal C}_{\mu\nu}^{ab} & = &
\G_{\Omega^a_\mu A^{*b}_\nu} 
- { i} \partial_\mu \partial_\nu D^{ab}
- { i} \int\!\diff^4z\left[
\partial_\mu D^{ad}\,{\widehat{\cal D}}_\rho^{d e}
\G_{\Omega^\rho_e A^{*b}_\nu} - 
{\widehat{\cal D}}_\rho^{de} \G_{\Omega^a_\mu 
A^{*\rho}_e} \partial_\nu D^{db} \right] \nonumber \\
&& 
- { i}\int\!\diff^4 z\!\!\int\!\diff^4 w \, 
{\widehat{\cal D}}_{\rho}^{d e}
\G_{\Omega^a_\mu A^{*\rho}_e}
D^{d m}
{\widehat{\cal D}}_{\sigma}^{m n}
\G_{\Omega^{\sigma}_{n} A^{*b}_\nu},
\label{sec.3.n.2}
\end{eqnarray}
and, since, as already noticed, $D$ is fixed by inverting
Eq.~(\ref{sec.2.3}), the r.h.s. of the equation above
 depends only on $\G_{\Omega A^*}$.

Eq.~(\ref{sec.3.n.2}) generalizes
to a non-trivial background configuration the results
obtained in~\cite{Grassi:2004yq} at zero background field.
The determination of $\G_{\Omega A^*}$ can be
performed once ${\cal C}$ is known
by expanding in the relevant form factors all
Green's functions in Eq.~(\ref{sec.3.n.2}); notice however that, as we will see below,   additional background-dependent invariants arise in the presence of a non-trivial
background configuration as compared to the $\widehat{A}\neq0$ case.

\subsection{An explicit example: the instanton background}

Though the focus of this paper is on the general properties of the formalism, it is nevertheless instructive to carry out an explicit computation with a given background in order to highlight the differences with respect to the $\widehat{A}=0$ case, as well as to familiarize with the calculation of the auxiliary function $\G_{\Omega A^*}$ which represents a key object in the ensuing analysis.

There are many possible topologically non-trivial background configurations that are believed to affect the (IR) dynamics of QCD Green's functions, and have been isolated, through cooling, in thermalized lattice configurations: vortices, monopoles and instantons~\cite{Maas:2005qt}. In what follows we will concentrate on an $SU(2)$ instanton background configuration, which, in the singular gauge reads~\cite{Schafer:1996wv}
\be
A^a_\mu(x) =2 \aeta^{a\mu\nu} \frac{x^\nu}{x^2(x^2+\rho^2)},
\label{inst}
\ee
where $ \aeta$ are the 't Hooft symbols:
\be
\aeta^{a\mu\nu}=\epsilon_{a\mu\nu4}-\delta_{a\mu}\delta_{\nu4}+\delta_{a\nu}\delta_{\mu4}; \qquad \epsilon_{1234}=1.
\ee
Since all the calculations will be perform in (Euclidean) momentum space, we parametrize the Fourier transform of the instanton configuration as
\be
A^a_\mu({ p}) = \aeta_{a\mu\nu} { p}^\nu f({ p}),
\ee
where (in the singular gauge) one finds
\be
f(p)=-i\frac{4 \pi^2  p^\nu}{\rho p^3} 
\left[-\frac{2}{p\rho} + K_1(p\rho) - p\rho K'_1(p\rho) \right],
\label{profile}
\ee
with $K$ the modified Bessel functions of the second kind; in the IR one has then
\be
f(p)\propto\frac{1}{p^2}.
\ee

Notice that in taking the Fourier transform we will drop all factors $(2\pi)^4$,
and in the case of fields, we will denote their Fourier transform by the same symbols as the original fields but with a momentum argument,
as identified by the Latin letters $p,\ q,\ r$ etc.
In order to keep the algebra as simple as possible
we work here in $SU(2)$ Yang-Mills theory;
the technique presented can be however extended in a rather straightforward way to more general gauge groups.

Let us notice, before starting the actual calculation, that the presence of a non-trivial background evidently breaks the translational invariance of the theory; this in turn means that a two-point function will feature two independent momenta $p$ and $q$ with~$p+q\neq0$. On the lattice this is a common situation for translational invariance is broken by the finite volume even  when $\widehat{A}=0$. 
To avoid the proliferation of form factors, and to keep the calculation at a reasonably simple level, we will assume in what follows that translational invariance has been recovered. This is equivalent to assuming that an averaging of some sort over the instanton collective coordinates (position, size and color orientation) is carried out; on the lattice this would correspond to measuring the two-point function as the average on many different background configurations, in a similar fashion to the method developed in~\cite{Boucaud:2005gg} for numerically inverting the Faddeev-Popov operator without imposing translational invariance, but rather recovering it as an average over many configurations.

\subsubsection{Ghost two-point function}

\begin{table}
\begin{tabular}{c||c|c|c|c|}
& $\quad \delta^{cd}\partial_\mu\partial_\nu\G_{\Omega^\mu_c A^{*\nu}_d} \quad$ & $\quad f^{ckd}\partial_\mu  \widehat A^k_\nu \G_{\Omega^\mu_c A^{*\nu}_d}\quad$ & $\quad  f^{ckd}\widehat A^k_\nu \partial_\mu \G_{\Omega^\mu_c A^{*\nu}_d}\quad$ & $\quad  -f^{ckd}\widehat A^k_\mu \partial_\nu \G_{\Omega^\mu_c A^{*\nu}_d}\quad$\\
\hline
\hline
$\ C_T(p^2)\ $ & 0 & 0 & 0 & 0\\
\hline
$\ C_L(p^2)\ $ & $3p^2$  & 0 & 0 &0 \\
\hline
$\ C_1^\eta(p^2)\ $ & 0 & $-6i(p+q)^2$ & $-6i(p+q)\cdot p$ & $-6i(p+q)\cdot p$\\
\hline
$\ C_2^\eta(p^2)\ $ & 0  & $-2i\frac{[(p+q)\wedge p]^2}{p^2}$  & 0 & $-6i(p+q)\cdot p$\\
\hline
$\ C_3^\eta(p^2)\ $ & 0 & $6i\frac{[(p+q)\cdot p]^2}{p^2}$ & $6i(p+q)\cdot p$ & 0\\
\hline
$\ C_4^\eta(p^2)\ $ & 0 & $-2i\frac{[(p+q)\wedge p]^2}{p^2}$ & 0 & 0\\
\hline
$\ C_5^\eta(p^2)\ $ & 0 & $2i\frac{[(p+q)\wedge p]^2}{p^2}$ & 0 & 0\\
\hline
\end{tabular}
\caption{\label{t0and1}Contribution to the different form factors from the terms of~(\ref{cov.1}) of zero and first order in the background field. To get the total contribution to the corresponding form factor one needs to multiply all entries by $f(p+q)$, except the one corresponding to $C_L$.}
\end{table}

In the presence of the background~(\ref{inst}) the two point functions $\Gamma_{\bar c c}$ and $\Gamma_{\Omega A^*}$ admit the following decomposition
\bea
\G_{\bar c^a c^b}(p) &=& { -}\delta^{ab} p^2 F^{-1}(p^2), \label{ghost}\\
\G_{\Omega_\mu^a A^{*b}_\nu}(p) & = &
\delta^{ab}   T_{\mu\nu}(p) C_T(p^2)+  \delta^{ab}L_{\mu\nu}(p)C_L(p^2)
+ f^{abc} \aeta_{c\mu\nu} C_1^\eta(p^2) 
+ f^{abc} \aeta_{c\mu\rho} \frac{p_\rho p^\nu}{p^2} C^\eta_2(p^2)\nonumber \\
&+& f^{abc} \aeta_{c\nu\rho} \frac{p_\rho p^\mu}{p^2} C_3^\eta(p^2)+ \aeta_{a\mu\rho} \aeta_{b\nu\sigma} \frac{p^\rho p^\sigma}{p^2}
C^\eta_4(p^2) + \aeta_{a\nu\rho} \aeta_{b\mu\sigma} \frac{p^\rho p^\sigma}{p^2} C^\eta_5(p^2),
\label{aux}
\eea
where $T_{\mu\nu}(p)=g_{\mu\nu} - p_\mu p_\nu/p^2$ (respectively, $L_{\mu\nu}(p)=p_\mu p_\nu/p^2$) is the dimensionless transverse (respectively, longitudinal) projector.
The ghost propagator $D$ is given by
\bea
D^{ab}(p^2) = \frac{i}{p^2} F(p^2) \delta^{ab}.
\eea

We then need to study Eq.~(\ref{sec.2.3}) and see what contributions we get from each of the form factors.
In the Landau gauge $\partial^\mu\widehat{A}_\mu=0$,
and thus the third term in (\ref{sec.2.3}) vanishes.
If momentum conservation is imposed, the ghost two-point function is proportional to $\delta^{ab}$ only and thus one can color trace the r.h.s. of eq.(\ref{sec.2.3}), simplifying considerably the algebra. The only contributions
come from the box term, which represents the standard kinetic term for the ghost field, and from the last term 
of eq.(\ref{sec.2.3}), involving
two background covariant derivatives. 
For the latter we get
\be
{\widehat{\cal D}}^{ac}_\mu {\widehat{\cal D}}^{ad}_\nu  = 
\partial_\mu \partial_\nu \delta_{cd} +
f^{ckd} \partial_\mu \widehat A^k_\nu 
+ f^{ckd}(\widehat A^k_\nu \partial_\mu - \widehat A^k_\mu \partial_\nu)
+ \widehat A^k_\mu \widehat A^k_\nu \delta^{cd} - \widehat A^d_\mu \widehat A^c_\nu.
\label{cov.1}
\ee
For the terms involving no or one background field the contributions to the product
${\widehat{\cal D}}^{ac}_\mu {\widehat{\cal D}}^{ad}_\nu \G_{\Omega^\mu_c A^{*\nu}_d}$ are shown in Table~\ref{t0and1}
(the instanton carries momentum $p+q$).

\begin{table}
\begin{tabular}{c||c|c|}
& $\quad\widehat{A}^k_\mu \widehat{A}^k_\nu \delta^{cd}\G_{\Omega^\mu_c A^{*\nu}_d}\quad$ & $\quad  \widehat{A}^c_\mu \widehat{A}^d_\nu\G_{\Omega^\mu_c A^{*\nu}_d}\quad$\\
\hline
\hline
$\ C_T(p^2)\ $ & $-9r^2+3\frac{(r\wedge p)^2}{p^2}$ & $-3r^2+\frac{(r\wedge p)^2}{p^2}$ \\
\hline
$\ C_L(p^2)\ $ & $-\frac{(r\wedge p)^2}{p^2}$  & $-\frac{(r\wedge p)^2}{p^2}$ \\
\hline
$\ C_1^\eta(p^2)\ $ & 0 & $6r^2$ \\
\hline
$\ C_2^\eta(p^2)\ $ & 0  & $2\frac{(r\wedge p)^2}{p^2}$  \\
\hline
$\ C_3^\eta(p^2)\ $ & 0 & $-2\frac{(r\wedge p)^2}{p^2}$ \\
\hline
$\ C_4^\eta(p^2)\ $ & $-3r^2+\frac{(r\wedge p)^2}{p^2}$ & $-3r^2+5\frac{(r\wedge p)^2}{p^2}$\\
\hline
$\ C_5^\eta(p^2)\ $ & $-3r^2+\frac{(r\wedge p)^2}{p^2}$ & $-9\frac{(p\cdot r)^2}{r^2}$ \\
\hline
\end{tabular}
\caption{\label{t2}Contribution to the different form factors from the terms of~(\ref{cov.1}) of second order in the background field. For each term the integral $\int\!\diff^4r\, f^2(r)$ is understood.}
\end{table}

For terms involving two background field insertions, there is one extra complication. Let's write 
\be
\widehat{A}^c_\mu(x) = \int\!\diff^4r\, \widehat{A}^c_\mu(r) \mathrm{e}^{-ir\cdot x};\qquad  \widehat A^d_\nu(x)\int\!\diff^4r'\, \widehat{A}^c_\mu(r') \mathrm{e}^{-ir'\cdot x},
\ee
for the two background fields; then substituting in~(\ref{cov.1}) and using the $\delta$ for the total momentum conservation, one is left with a residual integration over $\diff^4r$. The results for these terms are then shown in Table~\ref{t2}.

Therefore we obtain the following equation for the
2-point ghost dressing function $F^{-1}(p^2)$:
\bea
F^{-1}(p^2) = 1  - C_L(p^2) - \frac{1}{3 p^2}\Sigma(p^2)
\label{ghost.inst}
\eea
where $\Sigma$ denotes the sum of the contributions
spanned by the $C_j^\eta$ factors given in Table~\ref{t0and1} and by the $C_T,C_L$ and $C_j^\eta$ factors in Table~\ref{t2}. The $\Sigma$ term in eq.(\ref{ghost.inst}) takes into account
the instanton corrections to the well-known relation
\be
F^{-1}(p^2) = 1 - C_L(p^2)
\ee
which holds in a trivial background~\cite{Grassi:2004yq,Kugo:1995km,Aguilar:2009pp}.

\subsubsection{Kugo-Ojima function}

A second interesting quantity to look at in the presence of a non-trivial background is the Kugo-Ojima function $u$~\cite{Kugo:1979gm} which, when neglecting possible contributions from intermediate massive states, is related 
 in the limit of small momenta
to the connected function ${\cal C}$ of Eq.~(\ref{sec.3.n.1}) through
\be
u^{ab}(p^2) \left .  =  \right ._{p \rightarrow 0} - \frac13T^{\mu\nu}(p){\cal C}_{\mu\nu}^{ab}(p).
\label{ko}
\ee

Now, ${\cal C}$ admits { a} tensor decomposition
 of the same form as its 1PI part $\Gamma_{\Omega A^*}$
in eq.~(\ref{aux}); in this case however its form factors have both a 1PI contribution furnished by the various $C$s of~(\ref{aux}), as well as a reducible contribution coming from the second term in~(\ref{Cconn}). The important point however is that only color space diagonal structures emerges from the contraction~(\ref{ko}), that is one still has a diagonal Kugo-Ojima function
\be
u^{ab}(p^2) =\delta^{ab}u(p^2),
\label{diag}
\ee
 which is required in order to establish the Kugo-Ojima
criterion within the asymptotic Fock space
without violating the Faddeev-Popov ghost charge
conservation.

Using~(\ref{diag}), we can then write
\bea
u(p^2)&=&- \frac19T^{\mu\nu}(p)\left[\Gamma_{\Omega^a_\mu A^{*a}_\nu}(p)-\G_{\Omega^a_\mu \bar c^d}(p)\frac{F(p^2)}{p^2} 
\G_{c^d A^{*a}_\nu}(p)\right].
\label{u}
\eea
In order to isolate the relevant form factors contributing to $u$ let us introduce the following form factor decomposition of the 1-PI functions $\G_{\Omega \bar c}$ and
$\G_{c A^*}$
\bea
\G_{\Omega^a_\mu \bar c^d}(p) &=& \delta^{ad} p_\mu X_1(p^2)
+ f^{adk} \aeta_{k\mu\rho} p^\rho X_2(p^2), \nonumber \\
\G_{c^d A^{*b}_\nu}(p) &=& 
\delta^{db} p_\nu Y_1(p^2) +
f^{dbk} \aeta_{k\nu\rho} p^\rho Y_2(p^2),
\label{ko.ff}
\eea
which, once inserted in Eq.~(\ref{u}) above gives
\be
u(p^2)=-C_T(p^2)-\frac13\left[C^\eta_4(p^2)+C^\eta_5(p^2)\right]-\frac23F(p^2)X_2(p^2) Y_2(p^2).
\label{ko-bck}
\ee 

\begin{table}[!t]
\begin{tabular}{c||c|c|}
& \quad $-\partial^\rho \G_{\Omega^a_\mu A^{*d}_\rho}$ \quad & \quad  $-f^{dkc} \widehat{A}^\rho_k \G_{\Omega^a_\mu A^{*c}_\rho}$ \quad \\
\hline
\hline
$\ C_T(p^2)\ $ & 0 & $ -f^{adk} \aeta_{k\mu\rho} p^\rho $ \\
\hline
$\ C_L(p^2)\ $ & $ i\delta^{ad}p_\mu$ & $0$ \\
\hline
$\ C^\eta_1(p^2)\ $ & $i f^{adk} \aeta_{k\mu\rho} p^\rho $ & $- 2 \delta^{ad} p_\mu -f^{adk} \aeta_{k\mu\rho} p^\rho $\\
\hline
$\ C^\eta_2(p^2)\ $ & $i f^{adk} \aeta_{k\mu\rho} p^\rho $ &  $0$\\
\hline
$\ C^\eta_3(p^2)\ $ & $0$ & $2 \delta^{ad} p_\mu$ \\
\hline
$\ C^\eta_4(p^2)\ $ & $0$ & $0$ \\
\hline
$\ C^\eta_5(p^2)\ $ & $0$ & $f^{adk} \aeta_{d\mu\rho} p^\rho $ \\
\hline
\end{tabular}\hspace{.7cm}
\begin{tabular}{c||c|c|}
& \quad $\partial^\rho \G_{\Omega^d_\rho A^{*b}_\nu}$ \quad & \quad  $f^{dkc} \widehat{A}^\rho_k \G_{\Omega^c_\rho A^{*b}_\nu}$ \quad \\
\hline
\hline
$\ C_T(p^2)\ $ & 0 & $ f^{dbk} \aeta_{k\nu\rho} p^\rho $ \\
\hline
$\ C_L(p^2)\ $ & $ i\delta^{db}p_\nu$ & $0$ \\
\hline
$\ C^\eta_1(p^2)\ $ & $-i f^{dbk} \eta_{k\nu\rho} p^\rho $ & $- 2 \delta^{db} p_\nu +f^{dbk} \aeta_{k\nu\rho} p^\rho $\\
\hline
$\ C^\eta_2(p^2)\ $ & $0$ &  $- 2 \delta^{db} p_\nu$\\
\hline
$\ C^\eta_3(p^2)\ $ & $i f^{dbk} \eta_{k\nu\rho} p^\rho $ & $0$ \\
\hline
$\ C^\eta_4(p^2)\ $ & $0$ & $0$ \\
\hline
$\ C^\eta_5(p^2)\ $ & $0$ & $-f^{dbk} \aeta_{k\nu\rho} p^\rho $ \\
\hline
\end{tabular}
\caption{Contribution to the two-point functions $\Gamma_{\Omega\bar c}$ (left table) and $\Gamma_{cA^*}$ (right table) coming from the various $\Gamma_{\Omega A^*}$ form factors when using the relations~(\ref{age.2}) and~(\ref{sec.2.2}). In the second column of both tables multiplication by the instanton profile function $f(p^2)$ is understood.}
\label{t3}
\end{table}
The form factors $X_i$ and $Y_i$ can be then computed in terms
of the form factors of $\G_{\Omega A^*}$ alone by
using Eqs.~(\ref{age.2}) and~(\ref{sec.2.2}). Using the results reported in Tables~\ref{t3} we obtain\bea
X_1(p^2)&=&i\left[1+C_L(p^2)\right]-2f(p^2)\left[C_1^\eta(p^2)-C_3^\eta(p^2)\right],\nonumber \\
X_2(p^2)&=&i\left[C_1^\eta(p^2)+C_2^\eta(p^2)\right]+f(p^2)\left[C_5^\eta(p^2)-C_T(p^2)-C_1^\eta(p^2)\right],\nonumber \\
Y_1(p^2)&=&i\left[1+C_L(p^2)\right]-2f(p^2)\left[C_1^\eta(p^2)+C_2^\eta(p^2)\right],\nonumber\\
Y_2(p^2)&=&i\left[C_3^\eta(p^2)-C_1^\eta(p^2)\right]+f(p^2)\left[C_T(p^2)+C_1^\eta(p^2)-C_5^\eta(p^2)\right],
\eea
which once substituted in~(\ref{ko-bck}) provide the
 single instanton corrections to the relation $u=-C_T$ valid in the trivial background case~\cite{Grassi:2004yq,Kugo:1995km,Aguilar:2009pp}.

One should notice that a realistic
computation in the low energy regime requires to properly take into account the effects due to the overlapping of neighboring instantons~\cite{Schafer:1996wv};  in turn this implies not only that an appropriate average over the instanton collective coordinates is needed, but also that a modification of the profile function~(\ref{profile}) is mandatory, since the latter is valid only in the zero density limit.
A detailed discussion of this issue lies outside the scope of this paper, and therefore will not be pursued here.

\section{Gauge field redefinition at the quantum level\label{gfred}}

In the previous section we have shown that in the background
Landau gauge the behavior of the 2-point
1-PI ghost function at non-vanishing
background field $\widehat{A}\neq0$ 
is controlled by the function $\G_{\Omega A^*}$. Here we will show that
in the ghost-free sector, the very same functional
 $\G_{\Omega A^*}$ at $(A, \widehat{A}) \neq 0$ encodes the deformation of the background-quantum splitting induced by quantum corrections. 
This means that in the ghost-free sector one can obtain the 1-PI background-dependent amplitudes by performing
a certain (background-dependent) field redefinition, controlled by $\G_{\Omega A^*}$, on the 1-PI amplitudes involving only quantum fields.
Even though we will carry out the explicit calculations in the background Landau gauge, all results can be easily generalized to any other background gauge.

A few comments are in order here. 
It should be noticed that the WT identity of Eq.~(\ref{wti}) 
does not fix uniquely the dependence on the
background field $\widehat A$. For instance, in the space
of local functionals the most general solution
to Eq.~(\ref{wti}) is given by an arbitrary gauge invariant
functional constructed from the background field
strength $\widehat{F}^a_{\mu\nu} = 
\partial_\mu \widehat{A}^{a}_{\nu} - 
\partial_\nu \widehat{A}^{a}_{\mu} +  f^{abc} \widehat{A}^{b}_{\mu} \widehat{A}^{c}_\nu$, the fields $\chi$ transforming as matter fields
in the adjoint representation, and covariant
derivatives with respect to $\widehat{A}$.
However, already in perturbation theory it is known
that the actual dependence of $\fg$ on the background
field is much more constrained as a consequence of
the ST identity (\ref{me-1}). 
For instance, the term $(\widehat F^{a}_{\mu\nu})^2$ is allowed
by the WT identity, but it violates the ST identity.

As we will see below, this argument can be generalized. In the ghost-free sector $\Omega  = c= 0$
the ST identity can be solved in order to fix
uniquely the dependence on $\widehat{A}$
in terms of Green's functions that do not involve
background insertions. In this way one obtains
a formula for the background-quantum deformation
valid both in the full quantum theory as well as in a
non-perturbative setting, provided that the ST
identity (\ref{me-1}) is fulfilled.

In order to study this deformation, let us differentiate
the ST identity~(\ref{me-1}) 
with respect to $\Omega^a_\mu$ and finally set $(\Omega,c)=0$ 
while keeping both $A$ and $\widehat{A}$ different than zero; we find
\be
\G_{\widehat{A}^a_\mu}(x) = -\int\!\mathrm{d}^4y\left[\G_{\Omega^a_\mu A^{*\nu}_b }(x,y)\G_{A^b_\nu}(y)+b^b(y)\G_{\Omega_\mu^a \bar c^b}(x,y)\right].
\label{op-eq-1.n1}
\ee
Since no confusion can arise, in this section we set (with a slight abuse
of notation)
\be
\G_{\Omega^a_\mu A^{*b}_\nu\phi_1\cdots\phi_n} = 
\left . 
\frac{\delta^{(2+n)} \fg}{\delta \Omega^a_\mu \delta A^{*b}_\nu\delta\phi_1\cdots\delta\phi_n}
\right |_{(A,\widehat{A}) \neq 0}\hspace{-1.3cm}{\scriptsize \matrix{\vspace{-.2cm}(\phi^*,c,\Omega)=0}}
\label{g.astar.omega}
\ee

Accordingly we will explicitly display whenever Green's functions
are evaluated at zero background and quantum gauge fields.
In the ghost-free sector, Eq.~(\ref{op-eq-1.n1}) can be integrated explicitly. For that purpose it is convenient to introduce the reduced functional
$\rG$
\be
\rG=\G-\int\!\mathrm{d}^4x\,b^a(\widehat{\cal D}^\mu Q_\mu)^a.
\ee
This allows to take into account the $b$-dependence which is
confined at tree-level by the $b$-equation~(\ref{beq}).
Since
\be
\rG_{\widehat{A}^a_\mu}=\G_{\widehat{A}^a_\mu}-({\cal D}_\mu b)^a;\qquad
\rG_{A^a_\mu}=\G_{A^a_\mu}+(\widehat{\cal D}_\mu b)^a, 
\label{res-1}
\ee
one gets
\bea
\G_{\widehat{A}^a_\mu}(x)&=&-\int\!\mathrm{d}^4y\,\left\{\rG_{\Omega^a_\mu A^{*b}_\nu}(x,y)\left[\rG_{A^b_\nu}(y)-\widehat{\cal D}^{bd}_\nu b^d(y)\right]+b^b(y)\rG_{\Omega^a_\mu\bar c^b}(x,y)\right\}\nonumber \\
&=&-\int\!\mathrm{d}^4y\,\left\{\rG_{\Omega^a_\mu A^{*b}_\nu}(x,y)\rG_{A^b_\nu}(y)+\left[(\widehat{\cal D}^{bc}_\nu\rG_{\Omega^a_\mu A^{*c}_\nu}(x,y)+\rG_{\Omega^a_\mu\bar c^b}(x,y)\right]b^b(y)\right\}.\hspace{1cm}
\eea

On the other hand, from Eq.~(\ref{FPE}) we see that
\bea
\G_{\Omega^a_\mu \bar c^b}(x,y)&=&-\widehat{\cal D}^{bc}_\nu\G_{\Omega_a^\nu A^{*b}_\mu}(x,y)+{\cal D}^{ab}_\mu\delta(x-y),
\label{res.fpe.1}
\eea
Using Eq.~(\ref{res-1}) in conjunction with the last of Eq.~(\ref{res.fpe.1}), one then finds that the term ${\cal D}_\mu b$ drops out, leaving us with the 
simpler equation for $\rG$
\be
\rG_{\widehat{A}^a_\mu}(x)=-\int\!{\mathrm d}^4y\,\rG_{\Omega^a_\mu A^{*b}_\nu}(x,y)\rG_{\widehat{A}^b_\nu}(y).
\label{final}
\ee

Let us study Eq.~(\ref{final}) in the space of formal power series
in $\hat A,A$.
In order to find a solution to this equation, let us define a functional $\g$ such that
\be
\frac{\delta \g^b_\nu(y)}{\delta\widehat{A}^a_\mu(x)}=\G_{\Omega_\mu^a A^{*b}_\nu}(x,y).
\label{G.d.eq}
\ee
(Notice that due to the $b$ equation~(\ref{beq}) one has $\G_{\Omega A^*}\equiv\widetilde{\G}_{\Omega A^*}$). 

If we take the functional $\rG[\widehat{A},A]$, we see that the functional $\rG[0,A-{\cal G}]$ is then the sought for solution of Eq.~(\ref{final}), since
\bea
\rG_{\widehat{A}^a_\mu}(x)&=&-\int\!{\mathrm d}^4y\,\rG_{\widehat{A}^b_\nu}(y)\frac{\delta \g^b_\nu(y)}{\delta\widehat{A}^a_\mu(x)}\nonumber \\
&=&-\int\!{\mathrm d}^4y\,\G_{\Omega^a_\mu A^{*b}_\nu}(x,y)\rG_{\widehat{A}^b_\nu}(y).
\label{bkg.e.m}
\eea

Therefore, in the  sector where no ghost fields and no ghost external sources are present, the Slavnov-Taylor identity
entails that the full dependence on the background field is
generated by a redefinition of the 
quantum gauge field of the form
\be
A^a_\mu(x)\to A'^a_\mu(x)=A^a_\mu(x)-\g^a_\mu(x).
\label{repl}
\ee
with $\g^a_\mu$ solving the functional differential equation
(\ref{G.d.eq}). At tree-level this prescription is obviously correct since in the sector with no ghost fields and no ghost external sources there is no dependence on $A^a_\mu$ in the tree-level vertex functional $\fg^{(0)}$~(\ref{tlvf}) evaluated at $b^a = 0$, and $\G^{(0)}_{\Omega A^*}=0$. At the quantum level, however, $\g^a_\mu$ is non-trivial, and in particular it is non-linear in the background field. 

In order to illustrate
this point we explicitly construct $\g^a_\mu$ up to the second order term in the background field $\hat A$.
To lowest order, that is in the linear approximation,
Eq.~(\ref{G.d.eq}) has solution
\bea
\g^b_\nu(y)&=&
\int\!\diff^4z\, \G^{(A,\widehat{A})=0}_{\Omega_\mu^a A^{*b}_\nu}(z,y)
\widehat{A}^\mu_a(z) \, .
\label{lin.app}
\eea
Let us now expand $\G_{\Omega^a_\mu A^{*b}_\nu}$ up to 
first order in $A$ and $\widehat{A}$, obtaining
\bea
\G_{\Omega^a_\mu A^{*b}_\nu}(x,y)&=&
 \G_{\Omega^a_\mu A^{*b}_\nu} ^{\zb} (x,y)
+ \int\!\diff^4z\,
\G_{\Omega^a_\mu A^{*b}_\nu {\hat A}^c_\rho}^{\zb}
(x,y,z)\widehat{A}^\rho_c(z) \nonumber \\
&+&\int\!\diff^4z\,\G_{\Omega^a_\mu A^{*b}_\nu A^c_\rho}^{\zb}
(x,y,z)A^\rho_c(z)
+ \dots
\label{g.1}
\eea
Next, we replace the field $A^\rho_c$ appearing in the last term of the equation above by using the first order solution~(\ref{repl}), to get
\bea
\G_{\Omega^a_\mu A^{*b}_\nu}(x,y)&=&
\G_{\Omega^a_\mu A^{*b}_\nu}^{\zb}(x,y)
+ \int\!\diff^4z\,\G_{\Omega^a_\mu A^{*b}_\nu {\hat A}^c_\rho}^{\zb} (x,y,z)\widehat{A}^\rho_c(z)\nonumber\\
&+& \int\!\diff^4z\!\int\!\diff^4w\,
 \G_{\Omega^a_\mu A^{*b}_\nu A^c_\rho}^{\zb} (x,y,z)\,
\G_{\Omega^d_\sigma A^\rho_c}^{\zb} (w,z)\widehat{A}^\sigma_d(w)
+ \dots
\label{g.2}
\eea 
This last equation allows us to integrate Eq.~(\ref{G.d.eq}) up to second order in the background field, obtaining
\bea
\g^b_\nu(y)&=&\int\!\diff^4z\,  
\G_{\Omega_\mu^a A^{*b}_\nu}
^{\zb}
(z,y)\widehat{A}^\mu_a(z)
+\frac12\int\!\diff^4z_1\!\int\!\diff^4z_2\,
  \G_{\widehat{A}^{r_1}_{\rho_1}\Omega^{r_2}_{\rho_2} A^{*b}_\nu} ^{\zb} (z_1,z_2,y)
\widehat{A}^{\rho_1}_{r_1}(z_1)\widehat{A}^{\rho_2}_{r_2}(z_2)\nonumber \\
&+&\frac12\int\!\diff^4z_1\!\int\!\diff^4z_2\!\int\!\diff^4w\,
\G_{\Omega^{r_1}_{\rho_1} A^{*d}_\sigma}
^{\zb} (z_1,w)\,
\G_{A_{d}^{\sigma}\Omega^{r_2}_{\rho_2} A^{*b}_\nu}
^{\zb} (w,z_2,y)\widehat{A}^{\rho_1}_{r_1}(z_1)\widehat{A}^{\rho_2}_{r_2}(z_2).
\eea

We can easily check this result, by first differentiating the above equation with respect to the background field 
$\widehat{A}$ to get (for convenience, we suppress the space-time dependence)
\bea
\frac{\delta \g^b_\nu}{\delta\widehat{A}^a_\mu}&=&
\G_{\Omega^a_\mu A^{*b}_\nu}
^{\zb}
+\frac12
\int
\G_{\widehat{A}^{a}_{\mu}\Omega^{r_1}_{\rho_1} A^{*b}_\nu}
^{\zb}
\widehat{A}^{\rho_1}_{r_1}+
\frac12\int\!\!\!\!\int
\G_{\Omega^{a}_{\mu} A^{*d}_\sigma} ^{\zb}
\G_{A_{d}^{\sigma}\Omega^{r_1}_{\rho_1} A^{*b}_\nu}
^{\zb}
\widehat{A}^{\rho_1}_{r_1}\nonumber \\
&+&\frac12\int
\G_{\widehat{A}^{r_1}_{\rho_1}\Omega^{a}_{\mu} A^{*b}_\nu}
^{\zb}
\widehat{A}^{\rho_1}_{r_1}+
\frac12\int\!\!\!\!\int
\G_{\Omega^{r_1}_{\rho_1} A^{*d}_\sigma} 
^{\zb}
\G_{A_{d}^{\sigma}\Omega^{a}_{\mu} A^{*b}_\nu}
^{\zb}
\widehat{A}^{\rho_1}_{r_1}.
\label{gdiff}
\eea
Next, let us differentiate the functional form of the ST identity~(\ref{me-1})
with respect to two background sources $\Omega$ and one gluon anti-field $A^*$, obtaining (after setting fields and external sources to zero) the identity
\be
  \G_{\widehat{A}^a_\mu \Omega^b_\nu A^{*r}_{\rho}} 
^{\zb}
+
\int\!
\G_{\Omega^a_\mu A^{*d}_\sigma} 
^{\zb}\
\G_{A_d^\sigma\Omega^b_\nu A^{*r}_\rho} 
^{\zb} = \G_{\widehat{A}^b_\nu \Omega^a_\mu A^{*r}_{\rho}} 
^{\zb} +
\int\!
\G_{\Omega^b_\nu A^{*d}_{\sigma}} 
^{\zb}\
\G_{A_d^\sigma\Omega^a_\mu A^{*r}_\rho}
^{\zb},
\label{st.ggastar}
\ee
which, once inserted into  Eq.~(\ref{gdiff}), shows that it correctly reduces to Eq.~(\ref{g.2}).

The substitution rule of Eq.~(\ref{repl}) provides a powerful way to recover the full dependence on the background field, which can be extended beyond perturbation theory (provided that the ST identity in functional form (\ref{me-1}) are preserved also non perturbatively).  The 1-PI functions with the insertion of one source $\Omega$ and one anti-field $A^*$  are the important quantities controlling the quantum deformation of the background-quantum splitting. Indeed~(\ref{repl}) encodes in a simple form rather involved diagrammatic cancellations between the quantum and the background amplitudes which hold as a consequence of the ST identities. In the Appendix  we will illustrate these cancellations on the examples of the three-point functions  $\widehat{A}A A$ and $\widehat{A}\widehat{A} A$.

\subsection{Physical interpretation}

Now, let us turn our attention to the physical interpretation of the above result. If one assumes analyticity in the background fields of the 1-PI vertex functional, then what we have discovered is that the computation of  Green's functions for the gluon fields
in a non-trivial background is reduced to the computation of the same amplitudes at zero background and to the evaluation of the functional
$\G_{\Omega A^*}$, which fixes the quantum-background replacement (\ref{repl}) when loop corrections are taken into account.

As an example, consider the gluon two-point function $\Gamma_{AA}$ and let's calculate exploiting the above result its first correction due to a non-trivial background. First and foremost observe that in a non-trivial background the gluon propagator might not be transverse at all.
In order to prove this result let us write the $b$-equation
(\ref{beq}) in the Landau gauge
 for the connected generating functional $W$
(we introduce for a generic field $\phi$ its coupled source $J_\phi$):
\be
-J_{b^a} = \partial^\mu \frac{\delta W}{\delta J_{A^a_\mu}}-\partial \widehat A^a  +
f^{abc} \widehat A^b _\mu\frac{\delta W}{\delta J_{A_c^\mu}}
\label{e2}
\ee

Next, taking a second derivative w.r.t $J_A$, one has the identity
\be
0 = \partial^\mu \frac{\delta^2 W}{\delta J_{A^a_\mu} \delta J_{A_{b\nu}}} 
+
f_{adc} \widehat A_{d\mu} \frac{\delta^2 W}{\delta J_{A_{c\mu}} \delta J_{A_{b\nu}} }.
\ee
In the usual perturbation theory around a trivial background $\widehat{A}=0$, the
second  term on the r.h.s. of Eq.~(\ref{e2}) above vanishes, whence the transversality of the gluon propagator.
However when $\widehat{A}\neq0$ the situation is different, since the second term does not vanish in general.

This said, let us go back to the computation of the background corrections. The first of such corrections comes from the three-point gluon 1-PI Green function $\G_{AAA}$ at zero external background. Indeed, by keeping the lowest order term~(\ref{lin.app})  in the replacement rule~(\ref{repl}), one gets 
the following contribution 
\bea
-\frac{1}{2!}\int\! \G_{A_\alpha^c A^a_\mu A^b_\nu}
\G^{(A,\widehat{A})=0}_{\Omega^d_\rho A^{*\alpha}_c}
\widehat{A}^\rho_d  A^a_\mu A^b_\nu.
\label{2pt.bkg.contr}
\eea
Clearly such correction can be studied non-perturbatively. Indeed, $\G^{(A,\hat A)=0}_{\Omega A^*}$ can be determined from the 1-PI part of the correlation function $\langle T (D_\mu \bar c)^a (D_\nu  c)^b\rangle$ of Eq.~(\ref{sec.3.n.2}) after setting both the quantum and the background fields to zero, as was done, {\it e.g.}, in~\cite{Aguilar:2009pp}. The missing ingredient is then, on top of the explicit form of the background configuration, the conventional three-point gluon vertex $\G_{AAA}$ which in principle, however, is accessible to lattice studies. 

Notice finally that there is an infinite tower of corrections of the type~(\ref{2pt.bkg.contr}), coming from both multileg quantum functions ($\Gamma_{AAAA}$, etc.) as well as the higher order terms in the replacement formula~(\ref{repl});  in general, the order of the terms to be retained for capturing
the physics one wishes to describe depends on the particular
background under consideration.

\section{Solving the recursion for the background amplitudes}

\noindent In this Section we obtain an integral representation
of the vertex functional $\fg$ by solving the ST identity
(\ref{me-1}) via cohomological techniques \cite{Barnich:2000zw,Quadri:2002nh}.
This representation allows to explicitly isolate
the dependence on the background field  $\widehat{A}$.

For that purpose let us introduce the operator $\ope$ through
\begin{eqnarray}
\ope = \int\! \mathrm{d}^4x \, \Omega^a_\mu(x)\, \delta_{\widehat{A}^\mu_a(x)}; \qquad \ope^2 = 0,
\end{eqnarray}
where the nilpotency condition is due to the fermionic character of the background source $\Omega$.  Notice that, in order to avoid notational clutter, in the rest of the proof we will suppress the coordinates dependence of the various quantities when not necessary.

The ST identity in functional form~(\ref{me-1}) can be rewritten as
\be
\ope\fg=\operhs;\qquad\operhs\equiv-\int\!\mathrm{d}^4x\left[\fg_{A^{*\mu}_a}\fg_{A^{a}_\mu}+\fg_{c^{*a}}\fg_{c^a}+b^a\fg_{\bar c^a}\right]
\label{me-rew}
\ee
The above equation shows that $\fg$ can be seen as a solution 
to an inhomogeneous linear functional equation involving the nilpotent
operator $\ope$. The rhs of this equation, i.e.
$\operhs$, fulfills a consistency condition which follows
from the nilpotency of $\ope$
\be
\ope\operhs=0.
\label{opeonrhs}
\ee 

Next we introduce the homotopy operator $\homo$ as
\be
\homo = \int\!\mathrm{d}^4x\, \widehat{A}^a_\mu\!
\int_0^1\!\!\mathrm{d}t\, 
\lambda_t \,\delta_{\Omega^\mu_a},
\ee
where $\lambda_t$ represents an operator which acts on functionals  of the type $F[\widehat{A}^a_\mu, \Omega^a_\mu; \Phi']$ ($\Phi'$ denoting fields
and external sources other than the background fields $\widehat{A}$ and 
their ghost partners $\Omega$) as follows: it rescales 
by a factor of $t$ the $\widehat{A}$ and $\Omega$ variables,
while it does not act on the other variables:
\be
\lambda_t F[\widehat{A}^a_\mu, \Omega^b_\nu; \Phi']=F[t\widehat{A}^a_\mu, t\Omega^b_\nu; \Phi'].
\ee
Observing that, due to the fermionic nature of the source $\Omega$, one has the relations
\bea
\ope\homo&=&\int\!\mathrm{d}^4x\,\Omega^a_\mu(x)\int_0^1\!\mathrm{d}t\,\lambda_t\,\delta_{\Omega^\mu_a(x)}
+\int\!\mathrm{d}^4x\,\Omega^a_\mu(x)\int\!\mathrm{d}^4y\,\widehat{A}^b_\nu(y)\int_0^1\!\mathrm{d}t\,\delta_{\widehat{A}^\mu_a(x)}\lambda_t\,\delta_{\Omega^\nu_b(y)}\nonumber \\
\homo\ope&=&\int\!\mathrm{d}^4x\,\widehat{A}^a_\mu(x)\int_0^1\!\mathrm{d}t\,\lambda_t\,\delta_{\widehat{A}^\mu_a(x)}
-\int\!\mathrm{d}^4x\,\widehat{A}^a_\mu(x)\int_0^1\!\mathrm{d}t\,\lambda_t\!\int\!\mathrm{d}^4y\,\Omega_\nu^b(y)
\delta_{\Omega^\mu_a(x)}\delta_{\widehat{A}^\nu_b(y)},
\eea
it is relatively straightforward to see that, when working within the functional space spanned by monomials in which either $\widehat{A}$ or $\Omega$ appear at least once, the anticommutator of the $\ope$ and $\homo$ operators coincides with the functional identity in this space:
\be
\left\{\ope,\homo\right\}=\mathbb{I}.
\ee

Then, since Eq.~(\ref{me-rew}) implies that $\left.\operhs\right|_{\Omega=0}=0$, we see that $\operhs$ belongs to the functional space introduced above, and therefore we can write, using the property~(\ref{opeonrhs}),
\be
\operhs=\left\{\ope,\homo\right\}\operhs=\ope\homo\operhs.
\ee
Then, and again from Eq.~(\ref{me-rew}), we find the identity
\be
\ope\left(\fg-\homo\operhs\right)=0
\ee
which has the general solution
\be
\fg=\homo\operhs+\ope\Y+\fg_0,
\label{gen-sol}
\ee
where $\Y$ has ghost charge gh$(\Y)=-1$ and $\fg_0$ [which should not be confused with the tree-level vertex functional $\fg^{(0)}$ of Eq.~(\ref{tlvf})] does contain neither $\widehat{A}$ nor $\Omega$.

In the zero background ghost sector $\Omega=0$, the $\omega \Y$ term in Eq.~(\ref{gen-sol}) drops out, and one is left with the result 
\bea
\G &=&\homo\operhs+\fg_0\nonumber \\
&=& - \left . \int\! \mathrm{d}^4x \,{\widehat{A}^a_\mu(x)}\!\int_0^1\!\!\mathrm{d}t\,\lambda_t\,\delta_{\Omega^\mu_a(x)}\!\int\!\mathrm{d}^4y
\left[\fg_{A^{*\nu}_b}(y)\fg_{A^{b}_\nu}(y)+\fg_{c^{*b}}(y)\fg_{c^b}(y)+b^b(y)\fg_{\bar c^b}(y)\right]\right|_{\Omega=0}
\nonumber \\
&+&\fg_0.
\label{int.rep.g}
\eea
Finally, if one is interested in the sector where ghosts are absent, the formula above further simplifies to
\bea
\left . 
\G \right |_{c=0}
&=&- \left . \int\! \mathrm{d}^4x \,{\widehat{A}^a_\mu(x)}\!\int_0^1\!\!\mathrm{d}t\,\lambda_t\!\int\!\mathrm{d}^4y
\left[\fg_{\Omega^\mu_aA^{*\nu}_b}(x,y) \fg_{A^{b}_\nu}(y)
+b^b(y)\fg_{\Omega^\mu_a\bar c^b}(x,y)\right] \right|_{\Omega,c=0}\nonumber \\
&+&\left . \fg_0 \right |_{c=0}.
\label{me-final}
\eea
The equation above is quite remarkable,  for it provides a representation of 
the vertex functional in the ghost-free sector that isolates
the dependence on the background gauge field~$\widehat{A}$.

One can check that $\G$ in Eq.~(\ref{int.rep.g}) 
satisfies Eq.~(\ref{op-eq-1.n1}).
This will be established in two steps, by first checking that the $b$-equation is satisfied for then passing
to the equation of motion for the background field.

\subsection{$b$-equation}

Differentiating Eq.~(\ref{me-final}) with respect to the $b$ fields and using the vanishing of the three-point functions $\Gamma_{b^a\Omega^b_\mu A^{*d}_\nu}$ and $\Gamma_{b^a\Omega^b_\mu \bar c^c}$ following from the $b$-equation~(\ref{beq}), we get
\bea
\fg_{b^a}(x)&=&-\int\!\mathrm{d}^4y\,\widehat{A}^b_\mu(x)\int_0^1\!\!\mathrm{d}t\,\lambda_t\!\int\!\mathrm{d}^4z\left[\fg_{\Omega^\mu_b A^{*\nu}_c}(y,z)\fg_{b^a A^c_\nu}(x,y)+\delta(x-y)\fg_{\Omega^\mu_b \bar c^a}(y,z)\right]\nonumber \\
&+&\fg_{0b^a}(x).
\label{parres}
\eea
On the other hand, from Eqs.~(\ref{beq}) and~(\ref{FPE}) we have the results
\bea
\fg_{0b^a}(x)&=&\partial^\mu A^a_\mu(x) \nonumber \\
\fg_{b^aA^b_\mu}(x,y)&=&\widehat{\cal D}^{ab}_\mu\delta(x-y)
\nonumber \\
\fg_{\Omega^a_\mu \bar c^b}(x,y)&=&-\widehat{\cal D}^{bc}_\nu\fg_{\Omega_a^\nu A^{*b}_\mu}(x,y)+{\cal D}^{ab}_\mu\delta(x-y),
\label{res}
\eea
which once substituted into~(\ref{parres}) yield the gauge condition
\be
\fg_{b^a}(x)=\partial^{\mu}(A^a_\mu-\widehat{A}^a_\mu)+gf^{abc}\widehat{A}^\mu_bA^c_\mu,
\ee
prescribed by Eq.~(\ref{beq}) (in the Landau gauge $\xi=0$).

\subsection{Equation of motion for the background field $\widehat{A}$}

Let us now study the equation of motion of the $\widehat{A}$ field. The differentiation of Eq.~(\ref{me-final}) with respect to the background field 
$\widehat{A}$ yields
\bea
\Gamma_{\widehat{A}^a_\mu}(x)&=&
\left . 
-\int_0^1\!\!\mathrm{d}t\,\lambda_t\!\int\!\mathrm{d}^4y\left[\fg_{\Omega^a_\mu A^{*\nu}_b}(x,y)\fg_{A^b_\nu}(y)+b^b(y)\fg_{\Omega_\mu^a \bar c^b}(x,y)\right]
\right |_{\Omega=0}
\nonumber \\
&-&
\left . 
\int\!\mathrm{d}^4y\widehat{A}^b_\nu(y)\delta_{\widehat{A}^a_\mu(x)}\int_0^1\!\!\mathrm{d}t\,\lambda_t\!\int\!\mathrm{d}^4z\left[
\fg_{\Omega^\nu_b A^{*\rho}_c}(y,z)\fg_{A^c_\rho}(z)+b^c(z)\fg_{\Omega^\nu_b \bar c^c}(y,z)\right]
\right |_{\Omega=0}
\!\!\!\!\!,\hspace{1cm}
\label{op-eq}
\eea
where we see the appearance in the second line of the combination $\widehat{A}\,\delta_{\widehat{A}}$ which resembles the counting operator for the background field. 

On the other hand it is not difficult to realize that for each $\widehat{A}$ monomial that can possibly appear, the combinatorial factors induced by the operator $\lambda_t$ and the corresponding integral over d$t$ are such that Eq.~(\ref{op-eq}) reduces to the simpler relation
\be
\Gamma_{\widehat{A}^a_\mu}(x) = 
\left . 
-\int\!\mathrm{d}^4y\left[\fg_{\Omega^a_\mu A^{*\nu}_b}(x,y)\fg_{A^b_\nu}(y)+b^b(y)\fg_{\Omega_\mu^a \bar c^b}(x,y)\right] \right |_{\Omega=0} .
\label{op-eq-1}
\ee
Specifically, if, say, $k\widehat{A}^n$ is present in the rhs of Eq.~(\ref{op-eq-1}), then there are two terms of Eq.~(\ref{op-eq}) that could possibly contribute to it: the one corresponding to the first line, which will furnish $k/n+1$, and the one corresponding to the second line, which will give $k n/n+1$; therefore, the sum of the two contributions gives precisely the needed coefficient $k$. 

\section{Conclusions}

In this paper we have taken the first steps in developing  the  formal tools needed to solve the functional identities of Yang-Mills theories (Slavnov-Taylor, identities, $b$-equation, anti-ghost equation) in those cases where one has to deal with non-zero background configurations, such as the topologically non-trivial vacuum configurations provided by vortices, monopoles and  instantons. This is precisely what happens if one endeavors in implementing the BFM on the lattice, since in such case it has been shown long ago~\cite{Zwanziger:1982na} that a good background choice (that is one that fixes the gauge at least locally)  must be non partially flat in the sense of Eq.~(\ref{npf}),  which automatically excludes the trivial $\widehat{A}=0$ case. 

Our starting point has been the usual ST identities and the $b$-equation written in functional form, supplemented by the local anti-ghost identity and finally, in the background Landau gauge, the local ghost equation. We then first analyzed how the relations between 1PI gets modified by the presence of the background field, taking as an example the two-point ghost sector. Already at this level we saw the emergence of a fundamental quantity, namely the auxiliary function $\G_{\Omega A^*}$ which alone would determine completely the ghost two-point function even in the presence of a non-trivial background configuration.
Next, using as a toy background a single instanton configuration, we have calculated  the correction to both the ghost two-point function -- Eq.~(\ref{ghost.inst}) --, as well as to the Kugo-Ojima function -- Eq.~(\ref{ko-bck}) in terms of the form factors appearing in the Lorentz decomposition of $\G_{\Omega A^*}$.

In addition, when considering the ghost-free sector we were able to 
\begin{itemize}
\item Prove that the ST
identity can be solved in order to fix uniquely the dependence on $\widehat{A}$ in terms of Green's functions that do not involve background insertions. In this way one obtains the remarkable  formulas~(\ref{G.d.eq}) and~(\ref{repl}) for the background-quantum deformation valid both in the full quantum theory as well as in a non-perturbative setting, provided that the ST identity~(\ref{me}) is fulfilled;
\item Derive the representation of the vertex functional~(\ref{me-final}) that isolates the dependence on the background gauge field $\widehat{A}$.	
\end{itemize}

It should be noticed that, since $\G_{\Omega A^*}$ controls the quantum deformation of the classical background-quantum splitting in the zero ghost sector in a way compatible with the symmetries of the theory, one might reasonably conjecture that the full dependence
of the vertex functional on the background field (including the ghost-dependent sector)
can in fact be implemented via a canonical transformation (w.r.t the Batalin-Vilkovisky bracket of the model). In particular the approach based on canonical transformations might  be useful in order to obtain novel explicit representations of the 1-PI Green functions of Yang-Mills theory in the presence of a non-trivial background. Work along these directions is already in progress.

The techniques and results discussed here should be particularly useful in view of possible lattice implementation of the BFM, since they can be used as consistency check (if not as proper calculation tools) independently of the background chosen to calculate the correlation functions of interest. A second possible application would be in the calculation of Green's functions -- such as the gluon and ghost propagators -- through the corresponding Schwinger-Dyson equations in non-trivial backgrounds. Indeed, the analysis based on these latter equations presented so far in the literature, although accounting for the observed IR finiteness of the gluon propagator and the ghost-dressing functions and therefore in qualitative agreement with the lattice results~\cite{Aguilar:2008xm},  underestimate the size of both correlators. Rather than being due to the relevance of the diagrams left out (albeit in a gauge invariant fashion) in the truncation employed, an intriguing possibility is that this discrepancy might be related to the non-trivial structure of the vacuum, and in particular with the presence of topologically non-trivial configurations, such as vortices or monopoles; these configurations can be treated as a background and therefore accounted for through the techniques developed here,  as suggested by Eq.~(\ref{2pt.bkg.contr}) which provides the first correction to the (quantum) gluon two-point function due to a non-trivial background. 

Qualitative and quantitative comparisons with the effects observed on the lattice when removing center vortices from the vacuum configurations~\cite{de Forcrand:1999ms,Gattnar:2004bf} might at that point become possible.

\acknowledgements

We thank R. Ferrari for a critical reading of the manuscript, and 
acknowledge useful discussions with A. Cucchieri and J. Papavassiliou.

\appendix

\section{Perturbative analysis of the two-point sector}

In this appendix we discuss in some detail the perturbative two-point sector at zero background field, and in particular the renormalization of the auxiliary functions appearing in the expansion of the functional ${\cal G}$ of Section~\ref{gfred}.

Let us start by studying the gluon two-point functions. By keeping only the relevant terms in the rhs of (\ref{me-final}) and identifying term by term the lhs with the expression in the rhs, as explained in the previous example, one gets
\bea
\fg_{\widehat{A}^a_\mu A^b_\nu}(x,y)&=&-\int\!\mathrm{d}^4z\,\fg_{\Omega^a_\mu A^{*\rho}_c}(x,z)\fg_{A_\rho^cA^b_\nu}(z,y)\nonumber \\
\fg_{\widehat{A}^a_\mu\widehat{A}^b_\nu}(x,y)&=&-\int\!\mathrm{d}^4z\,\fg_{\Omega^a_\mu A^{*\rho}_c}(x,z)\fg_{A_\rho^c\widehat{A}^b_\nu}(z,y).
\eea

Next we perform the transformation $\widehat{A}\to\widehat{A}-Q$ and $A\to Q$ to get 
\bea
\fg_{\widehat{A}^a_\mu Q^b_\nu}(x,y)&=&\fg_{Q^a_\mu Q^b_\nu}(x,y)-\int\!\mathrm{d}^4z\,\fg_{\Omega^a_\mu A^{*\rho}_c}(x,z)\fg_{Q_\rho^c Q^b_\nu}(z,y)\nonumber \\
\fg_{\widehat{A}^a_\mu\widehat{A}^b_\nu}(x,y)&=&\fg_{Q^a_\mu\widehat{A}^b_\nu}(x,y)
-\int\!\mathrm{d}^4z\,\fg_{\Omega^a_\mu A^{*\rho}_c}(x,z)\fg_{Q_\rho^c\widehat{A}^b_\nu}(z,y),
\eea
where  in the second equation the dependence on the mixed background-quantum two point function has dropped out by using the first equation.  Taking the Fourier transform and setting the fields and sources to zero, one then recovers the usual background quantum identities of~\cite{Grassi:1999tp,Binosi:2002ez}, namely
\bea
\G_{Q^a_\mu \widehat{A}^b_\nu}(p)&=&[\delta^{br}g_\nu^\rho-\Gamma_{\Omega^b_\nu A^{*\rho}_r}(p)]\Gamma_{Q^a_\mu Q^r_\rho}(p)\nonumber \\
\G_{\widehat{A}^a_\mu \widehat{A}^b_\nu}(p)&=&[\delta^{br}g_\nu^\rho-\Gamma_{\Omega^b_\nu A^{*\rho}_r}(p)]\Gamma_{\widehat{A}^a_\mu Q^r_\rho}(p).
\label{bqis}
\eea

In the trivial background case $\widehat{A}=0$ Eq.~(\ref{aux}) reduces to
\be
\Gamma_{\Omega^a_\mu A^{*b}_\nu}(q)=\delta^{ab}[
T_{\mu\nu}(p) C_T(p^2) + L^{\mu\nu}(p) C_L(p^2)],
\label{decomp}
\ee
so that substituting the decomposition above into~(\ref{bqis}), and combining the resulting expressions, we obtain the relation
\be
\G_{\widehat{A}^a_\mu \widehat{A}^b_\nu}(p)=\left\{T^{\mu\rho}(p)[C_T(p^2)-1]^2+L^{\mu\rho}(p)[C_L(p^2)-1]^2\right\}\Gamma_{Q^a_\rho Q^b_\nu }(p).
\label{decomp.1}
\ee

By power-counting the divergence of $\G_{\Omega A^*}$ can only be proportional to $g_{\mu\nu}$; therefore, since the 1-PI functions do not have poles, the latter observation implies that the divergent parts of the transverse and longitudinal form factors  $C_T$ and $C_T$ (denoted by $\overline{C}_T$ and $\overline{C}_L$) are equal.

In addition, when setting to zero the background field $\widehat{A}$ Eq.~(\ref{sec.2.3}) reduces to 
\be
\Gamma_{c^a\bar c^b}(p)=p^2[1-C_L(p^2)]\delta^{ab}.
\label{ghid}
\ee

We can now discuss the renormalization of the auxiliary functions appearing in the functional ${\cal G}$, by exploiting the 
fact that the whole analysis above holds for the tree-level action plus counterterms, which we denote by $\overline{\G}$.

First of all notice that by power counting the functions $\Gamma_{\Omega A^* \phi_1\cdots\phi_n}$ are all superficially convergent so that we need to concentrate on $\Gamma_{\Omega A^*}$ only.
Next, observe that from Eq.~(\ref{decomp.1}) one gets (we suppress color indices)
\be
\overline{\G}^T_{\widehat{A}\widehat{A}}(p^2) = [C_T(p^2)-1]^2 \overline{\G}^T_{QQ}(p^2);
\qquad \overline{\G}^L_{\widehat{A}\widehat{A}}(p^2)= [C_L(p^2)-1]^2 \overline{\G}^L_{QQ}(p^2).
\label{cts}
\ee
We then set
\be
Z_{\widehat{A}}= \left . \frac{d}{dp^2} \overline{\G}^T_{\widehat{A}\widehat{A}}(p^2)\right |_{p^2 = 0};\qquad
Z_{Q}= \left . \frac{d}{dp^2} \overline{\G}^T_{QQ}(p^2)\right |_{p^2 = 0};\qquad
Z_c = \left . \frac{d}{dp^2} \overline{\G}_{c\bar c}(p^2)
 \right |_{p^2 = 0}  = [C_L(0)-1]
\ee
and notice that in the $p \to 0$ limit 
$C_L$ and $C_T$ coincide; in addition, by power-counting one can easily realize that the divergent part of $C_L$ and $C_T$
is a constant (no momentum dependence).
Then by differentiating Eq.~(\ref{cts})
with respect to $p^2$ and finally setting to $p^2 = 0$
one finds
\begin{eqnarray}
Z_{\widehat{A}}&=& \left . \frac{d}{dp^2} \overline{\G}^T_{\widehat{A}\widehat{A}}(p^2)\right |_{p^2 = 0}\nonumber \\
&  = & \left . \frac{d}{dp^2} [\overline{C}_T(p^2)-1]^2 \right |_{p^2=0}
\overline{\G}^T_{QQ}(0) +
[\overline{C}_T(0)-1]^2 \left . \frac{d}{dp^2}  \overline{\G}^T_{QQ}(p^2)
\right |_{p^2 = 0} \nonumber \\
& = & 
[\overline{C}_L(0)-1]^2 \left . \frac{d}{dp^2}  \overline{\G}^T_{QQ}(p^2)
\right |_{p^2 = 0} \nonumber \\
&=& Z_c^2 Z_Q.
\label{ct.3}
\end{eqnarray}

Indeed, since on general theoretical grounds identities like~(\ref{bqis}) are not deformed by the renormalization process, one has that $\Gamma_{\Omega A^*}$ renormalizes like $Z^{\frac12}_{\widehat{A}} Z_Q^{-\frac12}$; on the other hand, in the  Landau gauge Eq.~(\ref{ghid}) shows that $\Gamma_{\Omega A^*}$ renormalizes like $Z_c$~\cite{Aguilar:2009nf}. Eq.~(\ref{ct.3}) ensures the compatibility of the two renormalization conditions, and can be easily check up to the two-loop level. To be sure, when $N_\mathrm{f}=0$ (pure gluodynamics) and  $d=4+2\epsilon$ one has~\cite{Abbott:1980hw,Pascual:zb}
\bea
Z^{(2)}_Q&=&1+\frac{\alpha_s}\pi\frac{C_A}8\left(-\frac{13}3+\xi\right)\frac1\epsilon+\left(\frac{\alpha_s}\pi\right)^2\left[\frac{C^2_A}{32}\left(-\frac{13}4-\frac{17}{12}\xi+\frac12\xi^2\right)\frac1{\epsilon^2}\right.\nonumber \\
&+&\left.\frac{C^2_A}{128}\left(-\frac{59}2+\frac{11}2\xi+\xi^2\right)\frac1\epsilon\right]
\nonumber \\
Z^{(2)}_c&=&1+\frac{\alpha_s}{\pi}\frac{C_A}{16}\left(-3+\xi\right)\frac1\epsilon+\left(\frac{\alpha_s}\pi\right)^2\left[\frac{C^2_A}{512}\left(-35+3\xi^2\right)\frac1{\epsilon^2}+\frac{C^2_A}{1536}\left(-95-3\xi\right)\frac1\epsilon\right]
\nonumber \\
Z^{(2)}_{\widehat{A}}&=&1-\frac{\alpha_s}\pi\frac{11C_A}{12}\frac1\epsilon-\left(\frac{\alpha_s}\pi\right)^2\frac{34 C_A^2}{96}\frac1\epsilon,
\eea
where $\alpha_s=g^2/4\pi$. It is then easy to show that in the $\xi=0$ case
\bea
Z_c^{(1)}&=&\frac12\left(Z_{\widehat{A}}^{(1)}-Z_Q^{(1)}\right),\\
Z_c^{(2)}&=&\frac12\left(Z_{\widehat{A}}^{(2)}-Z_Q^{(2)}\right)+\frac18\left[3 \left(Z^{(1)}_Q\right)^2-
2Z^{(1)}_{\widehat{A}}Z^{(1)}_Q-\left(Z_{\widehat{A}}^{(1)}\right)^2\right].
\eea

\section{Perturbative diagrammatic cancellations}

Let us finally sketch the diagrammatic cancellations between the quantum and the background amplitudes which hold as a consequence of the ST identity, and are encoded in the substitution rule~(\ref{repl}). The philosophy adopted will be the following: we will start from the functional for the STI written in the background field method and prove that if one lets $A\to A+\g$ all the terms involving background fields vanish and thus we recover the ST identity written in terms of the quantum fields alone. To avoid notational clutter we will suppress all space-time dependence and integrals; in addition all Green's functions will be evaluated at zero fields (quantum and background), and we will not indicate this.

\subsection{Two-point sector}

We start by considering the fairly simple case of the (gluon) two-point functions. Let us scrutinize the mixed $\widehat{A}^{\rho_1}_{r_1}A^{a_1}_{\mu_1}$ first. Of all the field monomials appearing in the background generating functional, only two can possibly contribute to this amplitude (upon the replacement~$A\to A+\g$):
\bea
\frac12\G_{AA}AA \ &\quad \to& \quad\G_{A^{a_1}_{\mu_1}A^b_\nu}\G_{\Omega^{r_1}_{\rho_1}A^{*\nu}_b}\nonumber\\
\G_{\widehat{A}A}\widehat{A}A \ &\quad \to& \quad\G_{\widehat{A}^{r_1}_{\rho_1}A^{a_1}_{\mu_1}}.
\label{hAA}
\eea
On the other hand, differentiating the STI~(\ref{op-eq-1}) with respect to $A$, and setting afterwards all external sources and fields to zero, we get the identity
\be
\G_{\widehat{A}^{r_1}_{\rho_1}A^{a_1}_{\mu_1}}=-\G_{\Omega^{r_1}_{\rho_1}A^{*b}_\nu}\G_{A^\nu_bA^{a_1}_{\mu_1}},
\label{bqi.hAA}
\ee
by virtue of which the two terms in (\ref{hAA}) cancel.

In the $\widehat{A}^{\rho_1}_{r_1}\widehat{A}^{\rho_2}_{r_2}$ sector one has instead the following contributions
\bea
\frac12\G_{AA}AA \ &\quad \to& \quad\frac12\G_{A^{a_1}_{\mu_1}A^{a_2}_{\mu_2}}\G_{\Omega^{r_1}_{\rho_1}A^{*\mu_1}_{a_1}}\G_{\Omega^{r_2}_{\rho_2}A^{*\mu_2}_{a_2}}\nonumber\\
\G_{\widehat{A}A}\widehat{A}A \ &\quad \to& \quad\G_{\widehat{A}^{r_1}_{\rho_1}A^{a_1}_{\mu_1}}\G_{\Omega^{r_2}_{\rho_2}A^{*\mu_1}_{a_1}}\nonumber \\
\frac12\G_{\widehat{A}\widehat{A}}\widehat{A}\widehat{A}\ &\quad \to&\quad\frac12 \G_{\widehat{A}_{\rho_1}^{r_1}\widehat{A}_{\rho_2}^{r_2}}.
\label{hAhA}
\eea
We next differentiate the STI Eq.~(\ref{op-eq-1}) with respect to a background field $\widehat{A}$ to get, after setting external sources and fields to zero, the identity
\be
\G_{\widehat{A}^{r_1}_{\rho_1}\widehat{A}^{r_2}_{\mu_2}}=-\G_{\Omega^{r_1}_{\rho_1}A^{*b}_\nu}\G_{A^\nu_b\widehat{A}^{r_2}_{\rho_2}}.
\ee
Substituting this result and the identity~(\ref{bqi.hAA}) into the two last terms of (\ref{hAhA}) we get
\bea
\G_{\widehat{A}A}\widehat{A}A \ &\quad \to& \quad\frac12\G_{\widehat{A}^{r_1}_{\rho_1}A^{a_1}_{\mu_1}}\G_{\Omega^{r_2}_{\rho_2}A^{*\mu_1}_{a_1}}-\frac12\G_{\Omega^{r_1}_{\rho_1}A^{*b}_\nu}\G_{A^\nu_bA^{a_1}_{\mu_1}}\G_{\Omega^{r_2}_{\rho_2}A^{*\mu_1}_{a_1}}\nonumber \\
\frac12\G_{\widehat{A}\widehat{A}}\widehat{A}\widehat{A}\ &\quad \to&\, -\frac12\G_{\Omega^{r_1}_{\rho_1}A^{*b}_\nu}\G_{A^\nu_b\widehat{A}^{r_2}_{\rho_2}}.
\eea
Then, recalling that the indices $\rho_i$ and $r_i$ of the background fields are contracted, we see that also the $\widehat{A}\widehat{A}$ amplitude vanishes.

\subsection{Three-point sector}

Let us move now to the more complicate case of the three-point (gluon) sector, and start from the $\widehat{A}_{r_1}^{\rho_1}A_{a_1}^{\mu_1} A_{a_2}^{\mu_2}$ amplitude.  Of all the possible fields monomials appearing in the background generating functional, there are only three possible terms that, after the replacement  $A\to A+\g$, can possibly contribute to it, namely
\bea
\frac12\G_{AA}AA \ &\quad \to& \quad \G_{A^b_\nu A^{a_2}_{\mu_2}}\G_{\Omega^{r_1}_{\rho_1}A^{*\nu}_b A^{a_1}_{\mu_1}}\nonumber \\
\frac1{3!}\G_{AAA}AAA  &\quad \to& \quad \frac12\G_{A^b_\nu A^{a_1}_{\mu_1}A^{a_2}_{\mu_2}}\G_{\Omega^{r_1}_{\rho_1}A^{*\nu}_b}\nonumber \\
\frac1{2}\G_{\widehat{A}AA}\widehat{A}AA  &\quad \to& \quad
\frac12\G_{\widehat{A}^{r_1}_{\rho_1}A^{a_1}_{\mu_1}A^{a_2}_{\mu_2}}
\label{hAAA}
\eea
Differentiation of the STI~(\ref{op-eq-1}) with respect to two $A$ fields, provides, upon setting  external sources and fields to zero, the identity
\bea
\G_{\widehat{A}^{r_1}_{\rho_1}A^{a_1}_{\mu_1} A^{a_2}_{\mu_2}}&=&
-\G_{\Omega^{r_1}_{\rho_1}A^{*b}_\nu}\G_{A^\nu_bA^{a_1}_{\mu_1}A^{a_2}_{\mu_2}}
-\G_{\Omega^{r_1}_{\rho_1}A^{*b}_\nu A^{a_1}_{\mu_1}}\G_{A^\nu_bA^{a_2}_{\mu_2}}-\G_{\Omega^{r_1}_{\rho_1}A^{*b}_\nu A^{a_2}_{\mu_2}}\G_{A^\nu_b A^{a_1}_{\mu_1}}.
\label{bqi.hAAA}
\eea
We can then substitute the identity above in the last term of Eq.~(\ref{hAAA}); taking into account that the indices  of the $A$ fields are contracted, we get
\bea
\frac1{2}\G_{\widehat{A}AA}\widehat{A}AA  &\quad \to& \quad-\frac12\G_{\Omega^{r_1}_{\rho_1}A^{*b}_\nu}\G_{A^\nu_bA^{a_1}_{\mu_1}A^{a_2}_{\mu_2}}-\G_{\Omega^{r_1}_{\rho_1}A^{*b}_\nu A^{a_1}_{\mu_1}}\G_{A^\nu_bA^{a_2}_{\mu_2}}.
\eea
Summing up all the terms, we thus see that the amplitude $\widehat{A}AA$ vanishes, as it should.

As a last example consider finally the $\widehat{A}_{r_1}^{\rho_1}\widehat{A}_{r_2}^{\rho_2} A_{a_1}^{\mu_1}$ amplitude. In this case there are four terms that, after the replacement $A\to A+\g$, will contribute to this amplitude, and specifically
\bea
\frac12\G_{AA}AA  &\quad \to& \quad \frac12\G_{A^{a_1}_{\mu_1}A^b_\nu}\G_{\widehat{A}^{r_1}_{\rho_1}\Omega^{r_2}_{\rho_2}A^{*\nu}_b}
%
+ \frac12\G_{A^{a_1}_{\mu_1}A^b_\nu}\G_{\Omega^{r_1}_{\rho_1}A^{*\mu_2}_{a_2}}\G_{\Omega^{r_2}_{\rho_2}A^{*\nu}_{b}A^{a_2}_{\mu_2}} 
\nonumber \\
\frac1{3!}\G_{AAA}AAA  &\quad \to& \quad \frac12\G_{A^{a_1}_{\mu_1}A^{a_2}_{\mu_2}A^{a_3}_{\mu_3}}\G_{\Omega^{r_1}_{\rho_1}A^{*\mu_1}_{a_1}}\G_{\Omega^{r_2}_{\rho_2}A^{*\mu_2}_{a_2}}\nonumber \\
\frac12\G_{\widehat{A}AA}\widehat{A}AA  &\quad \to& \quad \G_{\widehat{A}^{r_1}_{\rho_1}A^{a_2}_{\mu_2}A^{a_1}_{\mu_1}}\G_{\Omega^{r_2}_{\rho_2}A^{*\mu_2}_{a_2}} 
\nonumber \\
\frac12\G_{\widehat{A}\widehat{A}A}\widehat{A}\widehat{A}A  &\quad \to& \quad 
\frac12\G_{\widehat{A}^{r_1}_{\rho_1}\widehat{A}^{r_2}_{\rho_2}A^{a_1}_{\mu_1}}.
\label{hAhAA}
\eea 
Let us now differentiate the STI~(\ref{op-eq-1}) withe respect to $\widehat{A}$ and $A$; after setting the external sources and fields to zero, one gets the STI
\bea
\G_{\widehat{A}^{r_1}_{\rho_1}\widehat{A}^{r_2}_{\rho_2}A^{a_1}_{\mu_1}}&=&-\G_{\Omega^{r_2}_{\rho_2}A^{*b}_\nu}\G_{\widehat{A}^{r_1}_{\rho_1}A^{a_1}_{\mu_1}A^\nu_b }
-\G_{\widehat{A}^{r_1}_{\rho_1}\Omega^{r_2}_{\rho_2}A^{*b}_\nu}\G_{A^{a_1}_{\mu_1}A^\nu_b}
-\G_{\Omega^{r_2}_{\rho_2}A^{*b}_\nu A^{a_1}_{\mu_1}}\G_{\widehat{A}^{r_1}_{\rho_1}A^\nu_b}.
\label{bqi.hAhAA}
\eea

We now use this identity and the ones of Eqs.~(\ref{bqi.hAA}) and~(\ref{bqi.hAAA}) in the last two equations of~(\ref{hAhAA}) to get
\bea
\frac12\G_{\widehat{A}AA}\widehat{A}AA \ &\quad \to& - \, 
\G_{\Omega^{r_1}_{\rho_1}A^{*\nu}_bA^{a_2}_{\mu_2}}
\G_{A^b_\nu A^{a_1}_{\mu_1}}
\G_{\Omega^{r_2}_{\rho_2}A^{*\mu_2}_{a_2}}
%
-\G_{\Omega^{r_1}_{\rho_1}A^{*\nu}_bA^{a_1}_{\mu_1}}
\G_{A^b_\nu A^{a_2}_{\mu_2}}
\G_{\Omega^{r_2}_{\rho_2}A^{*\mu_2}_{a_2}}
\nonumber \\
& & \,-\G_{\Omega^{r_1}_{\rho_1}A^{*\nu}_{b}}\G_{A^{a_1}_{\mu_1}A^b_\nu A^{a_2}_{\mu_2}}\G_{\Omega^{r_2}_{\rho_2}A^{*\mu_2}_{a_2}}
\nonumber \\
\frac12\G_{\widehat{A}\widehat{A}A}\widehat{A}\widehat{A}A  &\quad \to& \, -\frac12\G_{A^{a_1}_{\mu_1}A^b_\nu}\G_{\widehat{A}^{r_1}_{\rho_1}\Omega^{r_2}_{\rho_2}A^{*\nu}_b}
%
+\frac12\G_{\Omega^{r_2}_{\rho_2}A^{*\nu}_b}\G_{\Omega^{r_1}_{\rho_1}A^{*\mu_2}_{a_2}A^{a_1}_{\mu_1}}\G_{A^{a_2}_{\mu_2}A^b_\nu}
\nonumber\\
& & \, +\frac12\G_{\Omega^{r_2}_{\rho_2}A^{*\nu}_b}\G_{\Omega^{r_1}_{\rho_1}A^{*\mu_2}_{a_2}A^{b}_{\nu}}\G_{A^{a_2}_{\mu_2}A^{a_1}_{\mu_1}}
%
+\frac12\G_{\Omega^{r_2}_{\rho_2}A^{*\nu}_b}\G_{\Omega^{r_1}_{\rho_1}A^{*\mu_2}_{a_2}}\G_{A^{a_1}_{\mu_1}A^{a_2}_{\mu_2}A^{b}_{\nu}}
\nonumber \\
& & \, +\frac12\G_{\Omega^{r_2}_{\rho_2}A^{*\nu}_bA^{a_1}_{\mu_1}}\G_{\Omega^{r_1}_{\rho_1}A^{*\mu_2}_{a_2}}\G_{A^{a_2}_{\mu_2}A^b_\nu }
\eea
Adding all together taking into account the contracted indices, we see that all the terms (and therefore the 
$\widehat{A}\widehat{A}A$ amplitude) vanish, according to the patterns shown in Table~\ref{t-hAhAA}.

\begin{table}[!t]
\begin{tabular}{c||c|c|c|c|c|}
& $\quad\frac12\G_{AA}AA\quad$ &
$\quad\frac1{3!}\G_{AAA}AAA\quad$ &
$\quad\frac12\G_{\widehat{A}AA}\widehat{A}AA\quad$ &
$\quad\frac12\G_{\widehat{A}\widehat{A}A}\widehat{A}\widehat{A}A\quad$ & $\quad$ Sum $\quad$\\
\hline\hline
$\propto\G_{AAA}$ & 0 & $1$ & $-2$ & $1$ & 0\\
\hline
$\propto\G_{A^b_\nu A^{a_1}_{\mu_1}}$ & $\frac12$ & 0 & $-1$ & $\frac12$ & 0\\
\hline
$\propto\G_{A^b_\nu A^{a_2}_{\mu_2}}$ & 0 & 0 & $-1$ & $1$ & 0 \\
\hline
$\propto\G_{\widehat{A}\Omega A^*}$ & $\frac12$ & 0 & 0 & $-\frac12$  & 0\\
\hline
\end{tabular}
\caption{\label{t-hAhAA} The cancellations between different terms contributing to the $\widehat{A}\widehat{A}A$ amplitude after implementing  the substitution rule~(\ref{repl}). Notice also that the second row refers to all terms proportional to $\G_{A^b_\nu A^{a_1}_{\mu_1}}$  which are not of the type $\G_{\widehat{A}\Omega A^*}$ (which explicitly appears in the last row).
}
\end{table}


\begin{thebibliography}{99}

\bibitem{DeWitt:1967ub}
  B.~S.~DeWitt,
  Phys.\ Rev.\  {\bf 162}, 1195 (1967);
  J.~Honerkamp,
  Nucl.\ Phys.\  {\bf B48}, 269 (1972);
  R.~E.~Kallosh,
  Nucl.\ Phys.\  {\bf B78}, 293 (1974);
  H.~Kluberg-Stern, J.~B.~Zuber,
  Phys.\ Rev.\  {\bf D12}, 482-488 (1975);
  I.~Y.~.Arefeva, L.~D.~Faddeev, A.~A.~Slavnov,
  Theor.\ Math.\ Phys.\  {\bf 21}, 1165 (1975);
G.~'t~Hooft, {The Background Field Method in Gauge Field Theories, }In *Karpacz
  1975, Proceedings, Acta Universitatis Wratislaviensis No.368, Vol.1*, Wroclaw
 345 (1976);
  S.~Weinberg,
  Phys.\ Lett.\  {\bf B91}, 51 (1980);
  G.~M.~Shore,
  Annals Phys.\  {\bf 137}, 262 (1981);
  L.~F.~Abbott, M.~T.~Grisaru, R.~K.~Schaefer,
  Nucl.\ Phys.\  {\bf B229}, 372 (1983);
  C.~F.~Hart,
  Phys.\ Rev.\  {\bf D28}, 1993-2006 (1983).

\bibitem{Abbott:1980hw}
L.~F.~Abbott,
  Nucl.\ Phys.\  B {\bf 185}, 189 (1981);
  Acta Phys.\ Polon.\  {\bf B13}, 33 (1982).

\bibitem{Becchi:1999ir}
  C.~Becchi, R.~Collina,
  Nucl.\ Phys.\  {\bf B562}, 412-430 (1999).

\bibitem{Ferrari:2000yp}
  R.~Ferrari, M.~Picariello, A.~Quadri,
  Annals Phys.\  {\bf 294}, 165-181 (2001).

\bibitem{Ichinose:1981uw}
  S.~Ichinose, M.~Omote,
  Nucl.\ Phys.\  {\bf B203}, 221 (1982);
  D.~M.~Capper, A.~MacLean,
  Nucl.\ Phys.\  {\bf B203}, 413 (1982).

\bibitem{Denner:1994xt}
  A.~Denner, G.~Weiglein, S.~Dittmaier,
  Nucl.\ Phys.\  {\bf B440}, 95-128 (1995).

\bibitem{Gates:1983nr}
  S.~J.~Gates, M.~T.~Grisaru, M.~Rocek {\it et al.},
  Front.\ Phys.\  {\bf 58}, 1-548 (1983).

\bibitem{Cornwall:1981zr}
J.~M.~Cornwall,
Phys.\ Rev.\ D {\bf 26}, 1453 (1982).


\bibitem{Cornwall:1989gv}
  J.~M.~Cornwall and J.~Papavassiliou,
  Phys.\ Rev.\  D {\bf 40}, 3474 (1989);
for a recent review on the subject see also 
  D.~Binosi, J.~Papavassiliou,
  Phys.\ Rept.\  {\bf 479}, 1-152 (2009).

\bibitem{Binosi:2002ft}
  D.~Binosi and J.~Papavassiliou,
  Phys.\ Rev.\ D {\bf 66}(R), 111901 (2002);
  J.\ Phys.\ G {\bf 30}, 203 (2004).

\bibitem{Binosi:2007pi}
  D.~Binosi and J.~Papavassiliou,
  Phys.\ Rev.\  D {\bf 77}(R), 061702 (2008);
  JHEP {\bf 0811}, 063 (2008).


\bibitem{Grassi:1999tp}
  P.~A.~Grassi, T.~Hurth and M.~Steinhauser,
  Annals Phys.\  {\bf 288}, 197 (2001);

\bibitem{Binosi:2002ez}
  D.~Binosi and J.~Papavassiliou,
  Phys.\ Rev.\ D {\bf 66}, 025024 (2002).


\bibitem{Aguilar:2006gr}
  A.~C.~Aguilar, J.~Papavassiliou,
  JHEP {\bf 0612}, 012 (2006).

\bibitem{Aguilar:2008xm}
  A.~C.~Aguilar, D.~Binosi, J.~Papavassiliou,
  Phys.\ Rev.\  {\bf D78}, 025010 (2008).

\bibitem{Cucchieri:2007md}
  A.~Cucchieri and T.~Mendes,
  PoS {\bf LAT2007}, 297 (2007);
  Phys.\ Rev.\ Lett.\  {\bf 100}, 241601 (2008).


\bibitem{Bogolubsky:2007ud}
  I.~L.~Bogolubsky, E.~M.~Ilgenfritz, M.~Muller-Preussker and A.~Sternbeck,
  PoS {\bf LAT2007}, 290 (2007);
  Phys.\ Lett.\  {\bf B676}, 69-73 (2009).

\bibitem{Kugo:1979gm}
  T.~Kugo and I.~Ojima,
  Prog.\ Theor.\ Phys.\ Suppl.\  {\bf 66}, 1 (1979).

\bibitem{Gribov:1977wm}
 V.~N.~Gribov,
  Nucl.\ Phys.\  B {\bf 139}, 1 (1978).

\bibitem{Zwanziger:1993dh}
  D.~Zwanziger,
  Nucl.\ Phys.\  B {\bf 412}, 657 (1994).


\bibitem{Dudal:2008sp}
For a successful attempt to reconcile the Gribov-Zwanzinger scenario with the lattice results see D.~Dudal, J.~A.~Gracey, S.~P.~Sorella, N.~Vandersickel and H.~Verschelde,
Phys.\ Rev.\  D {\bf 78}, 065047 (2008).

\bibitem{Cucchieri:2010xr}
 For a thorough and concise review of all these aspects of the lattice calculations, see  A.~Cucchieri and T.~Mendes,
  PoS {\bf QCD-TNT09}, 026 (2009).

\bibitem{Cucchieri:2011aa}
  A.~Cucchieri, T.~Mendes, G.~M.~Nakamura and E.~M.~S.~Santos,
  arXiv:1101.5080 [hep-lat];
  A.~Cucchieri, T.~Mendes and E.~M.~S.~Santos,
  Phys.\ Rev.\ Lett.\  {\bf 103}, 141602 (2009).

\bibitem{Dashen:1980vm}
In the Feynman gauge, a possible implementation of the BFM has been proposed long ago in R.~F.~Dashen and D.~J.~Gross,
  Phys.\ Rev.\  D {\bf 23}, 2340 (1981).

\bibitem{Zwanziger:1982na}
  D.~Zwanziger,
  Nucl.\ Phys.\  {\bf B209}, 336 (1982).

\bibitem{de Forcrand:1999ms}
  P.~de Forcrand and M.~D'Elia,
  Phys.\ Rev.\ Lett.\  {\bf 82}, 4582 (1999).

\bibitem{Gattnar:2004bf}
  J.~Gattnar, K.~Langfeld and H.~Reinhardt,
  Phys.\ Rev.\ Lett.\  {\bf 93}, 061601 (2004)
  [arXiv:hep-lat/0403011].

\bibitem{Batalin:1977pb}
  I.~A.~Batalin, G.~A.~Vilkovisky,
  Phys.\ Lett.\  {\bf B69}, 309-312 (1977);
  Phys.\ Lett.\  {\bf B102}, 27-31 (1981).



\bibitem{Gomis:1994he}
  See e.g. J.~Gomis, J.~Paris, S.~Samuel,
  Phys.\ Rept.\  {\bf 259 } (1995)  1-145.
  [hep-th/9412228].



\bibitem{Piguet:1995er}
  O.~Piguet, S.~P.~Sorella,
  Lect.\ Notes Phys.\  {\bf M28 } (1995)  1-134.

\bibitem{Barnich:2000zw}
  G.~Barnich, F.~Brandt, M.~Henneaux,
  Phys.\ Rept.\  {\bf 338 } (2000)  439-569.



\bibitem{Quadri:2002nh}
  A.~Quadri,
  JHEP {\bf 0205 } (2002)  051.

\bibitem{Barnich:1994ve}
  G.~Barnich, M.~Henneaux,
  Phys.\ Rev.\ Lett.\  {\bf 72 } (1994)  1588-1591.



\bibitem{Grassi:2004yq}
  P.~A.~Grassi, T.~Hurth,  A.~Quadri,
  Phys.\ Rev.\  {\bf D70}, 105014 (2004).

\bibitem{Maas:2005qt}
  A.~Maas,
  Eur.\ Phys.\ J.\  C {\bf 48}, 179 (2006);
  Nucl.\ Phys.\  A {\bf 790}, 566 (2007).

\bibitem{Schafer:1996wv}
  See, {\it e.g.}, T.~Schafer and E.~V.~Shuryak,
  Rev.\ Mod.\ Phys.\  {\bf 70}, 323 (1998), and references therein.

\bibitem{Boucaud:2005gg}
  P.~Boucaud {\it et al.},
  Phys.\ Rev.\  D {\bf 72}, 114503 (2005).

\bibitem{Kugo:1995km}
  T.~Kugo,
  arXiv:hep-th/9511033;

\bibitem{Aguilar:2009pp}
  A.~C.~Aguilar, D.~Binosi and J.~Papavassiliou,
  JHEP {\bf 0911}, 066 (2009).

\bibitem{Aguilar:2009nf}
  A.~C.~Aguilar, D.~Binosi, J.~Papavassiliou and J.~Rodriguez-Quintero,
  Phys.\ Rev.\  D {\bf 80}, 085018 (2009).

\bibitem{Pascual:zb}
P.~Pascual and R. Tarrach, {\it QCD: Renormalization for the Practitioner}, 
Springer and Verlag, Heidelberg (1984).

\end{thebibliography}
\end{document}